\documentclass[traditabstract]{aa}
\usepackage{txfonts}
\usepackage{natbib}
\usepackage{graphicx}
\usepackage{afterpage}
\usepackage{lscape}
\usepackage{psfrag}
\usepackage[hyphens]{url}
\usepackage{microtype}
\usepackage{float}
\usepackage{afterpage}
\usepackage{longtable}
\usepackage{fixltx2e}
\usepackage{multirow}
\makeatletter
\def\url@leostyle{%
  \@ifundefined{selectfont}{\def\UrlFont{\sf}}{\def\UrlFont{\small\ttfamily}}}
\makeatother
\def\apjl{ApJL}

\def\aap{A\&A}
\def\apjs{ApJS}

\newcommand{\footnoteremember}[2]{
\footnote{#2}
\newcounter{#1}
\setcounter{#1}{\value{footnote}}
}
\newcommand{\footnoterecall}[1]{
\footnotemark[\value{#1}]
}

\begin{document}

   \title{A resolved analysis of cold dust and gas in the nearby edge-on spiral NGC~891\thanks{Based on observations from {\it Herschel}, an ESA space observatory with science instruments provided by European-led Principal Investigator consortia and with important participation from NASA.}}

   \subtitle{ }

   \author{T. M. Hughes\inst{1},  M. Baes\inst{1}, J. Fritz\inst{1}, M. W. L. Smith\inst{2}, T. J. Parkin\inst{3}, G. Gentile\inst{1,4}, G. J. Bendo\inst{5}, C. D. Wilson\inst{3},\\ F. Allaert\inst{1}, S. Bianchi\inst{6}, I. De Looze\inst{1}, J. Verstappen\inst{1}, S. Viaene\inst{1}, M. Boquien\inst{7}, A. Boselli\inst{8}, D. L. Clements\inst{9},\\  J. I. Davies\inst{2}, M. Galametz\inst{10}, S. C. Madden\inst{11}, A. R\'emy-Ruyer\inst{11}, L. Spinoglio\inst{12}}
	  
\institute{Sterrenkundig Observatorium, Universiteit Gent, Krijgslaan 281-S9, Gent 9000, Belgium
                \email{thomas.hughes@ugent.be}
	      	\and
	      	School of Physics \& Astronomy, Cardiff University, The Parade, Cardiff CF24 3AA, UK	
			\and
                Department of Physics \& Astronomy, McMaster University, Hamilton, Ontario L8S 4M1, Canada 
             \and   
	      	Department of Physics and Astrophysics, Vrije Universiteit Brussel, Pleinlaan 2, 1050 Brussels, Belgium               
              \and
                UK ALMA Regional Centre Node, Jodrell Bank Centre for Astrophysics, School of Physics and Astronomy, \\ University of Manchester, Oxford Road, Manchester M13 9PL, UK                 
                 \and
                INAF - Osservatorio Astrofisico di Arcetri, Largo E. Fermi 5, 50125, Florence, Italy                  
                \and
                Institute of Astronomy, University of Cambridge, Madingley Road, Cambridge CB3 0HA, UK             
                \and
                Aix-Marseille Universit\'e, CNRS, d'Astrophysique de Marseille, UMR 7326, F-13388 Marseille, France
                \and
                Astrophysics Group, Imperial College London, Blackett Laboratory, Prince Consort Road, London SW7 2AZ, UK
                \and
                European Southern Observatory, Karl-Schwarzschild Str. 2, 85748 Garchin bei Muenchen, Germany 
                \and
                Laboratoire AIM, CEA/DSM-CNRS-Universite Paris Diderot DAPNIA/Service d'Astrophysique, \\ 
Bat. 709, CEA-Saclay, F-91191 Gif-sur-Yvette Cedex, France
                \and
                Istituto di Astrofisica e Planetologia Spaziali, INAF-IAPS, Via Fosso del Cavaliere 100, I-00133 Roma, Italy
                 } 

   \date{Accepted for publication in A\&A.}

\newcommand{\hi}{H{\sc i}} 
\newcommand{\hii}{H{\sc ii}\ }
\newcommand{\oi}{O{\sc i}}
\newcommand{\oii}{O{\sc ii}}
\newcommand{\hd}{H{\sc $\delta$}}
\newcommand{\hg}{H{\sc $\gamma$}}
\newcommand{\hb}{H{\sc $\beta$}}
\newcommand{\oiii}{O{\sc iii}}
\newcommand{\oiv}{O{\sc iv}}
\newcommand{\nii}{N{\sc ii}}
\newcommand{\niii}{N{\sc iii}}
\newcommand{\ha}{H{\sc $\alpha$}}
\newcommand{\sii}{S{\sc ii}}
\newcommand{\siii}{S{\sc iii}}
\newcommand{\cone}{C$_{1}$}
\newcommand{\rz}{R$_{23}$}
\newcommand{\oz}{O$_{32}$}
\newcommand{\zzz}{12+log(O/H)}
\newcommand{\kms}{km~s$^{-1}$\ }
\newcommand{\sdust}{$\Sigma_{\mathrm{dust}}$}
\newcommand{\sgas}{$\Sigma_{\mathrm{gas}}$}
\newcommand{\shii}{$\Sigma_{\mathrm{H}_{2}}$} 
\newcommand{\shi}{$\Sigma_{\mathrm{H}\tiny{\textsc{i}}}$} 
\newcommand{\ssfr}{$\Sigma_{\mathrm{SFR}}$}

  \abstract{We investigate the connection between dust and gas in the nearby edge-on spiral galaxy NGC~891, a target of the Very Nearby Galaxies Survey. High resolution \textit{Herschel} PACS and SPIRE 70, 100, 160, 250, 350, and 500~$\mu$m images are combined with JCMT SCUBA 850~$\mu$m observations to trace the far-infrared/submillimetre spectral energy distribution (SED). Maps of the \hi \ 21 cm line and CO(J=3-2) emission trace the atomic and molecular hydrogen gas, respectively. We fit one-component modified blackbody models to the integrated SED, finding a global dust mass of (8.5 $\pm$ 2.0) $\times$ 10$^{7}$ M$_{\odot}$ and an average temperature of 23 $\pm$ 2 K, consistent with results from previous far-infrared experiments. We also fit one-component modified blackbody models to pixel-by-pixel SEDs to produce maps of the dust mass and temperature. The dust mass distribution correlates with the total stellar population as traced by the 3.6~$\mu$m emission. The derived dust temperature, which ranges from approximately 17 to 24 K, is found to correlate with the 24~$\mu$m emission. Allowing the dust emissivity index to vary, we find an average value of $\beta$ = 1.9 $\pm$ 0.3. We confirm an inverse relation between the dust emissivity spectral index and dust temperature, but do not observe any variation of this relationship with vertical height from the mid-plane of the disk. A comparison of the dust properties with the gaseous components of the ISM reveals strong spatial correlations between the surface mass densities of dust (\sdust) and the molecular hydrogen (\shii) and total gas surface densities (\sgas). These observations reveal the presence of regions of dense, cold dust that are coincident with peaks in the gas distribution and are associated with a molecular ring. Furthermore, the observed asymmetries in the dust temperature, the H$_{2}$-to-dust ratio and the total gas-to-dust ratio hint that an enhancement in the star formation rate may be the result of larger quantities of molecular gas available to fuel star formation in the NE compared to the SW. Whilst the asymmetry likely arises from dust obscuration due to the geometry of the line-of-sight projection of the spiral arms, we cannot exclude that there is also an enhancement in the star formation rate in the NE part of the disk.}


	\authorrunning{T. M. Hughes et al.}
	\titlerunning{Resolved dust and gas in NGC~891}
   \maketitle
 
\section{Introduction}

Dust and gas are crucial ingredients for star formation in the Universe. Dust grains provide a surface on which molecular hydrogen (H$_{2}$) gas forms from atomic hydrogen (\hi) gas via catalytic reactions (\citealp*{gould1963}; \citealp*{hollenbach1971}; \citealp*{cazaux2002}). Regions of cooling and fragmentation of molecular gas in giant molecular clouds (GMCs) are the dominant formation sites of stars in galaxies (see \citealp*{fukui2010}, and references therein). These stars produce the heavy elements via nucleosynthesis in their cores. Upon the death of a star, metals are expelled into the interstellar medium (ISM) and either mix with the gas phase or condense to form dust grains in the metal-rich cooling gas (see e.g., \citealp*{nozawa2013}), such as in the expanding ejecta of novae and supernovae (e.g., \citealp{Kozasa1991}; \citealp{nozawa2003}; \citealp{barlow2010}), the stellar winds from cool, dense atmospheres of asymptotic giant branch stars (e.g., \citealp*{ferrarotti2006}; \citealp*{zhukovska2013}), and the common-envelope in close binary systems (e.g., \citealp{lu2013}). Growth, modification and destruction processes may affect the overall abundance of interstellar dust (e.g., \citealp{clemens2010}; \citealp{mattsson2012}), whilst surviving grains have the potential to aid future episodes of star formation. Thus, the evolutionary processes governing gas, stars, metals and dust are mutually related, and constraining the relationships between these four components is indispensable for understanding galaxy formation and evolution. 

Detailed studies of the interaction between gas and stars have been technologically feasible for a long time. Global star formation rates (SFRs) of galaxies are observed to be well correlated with their atomic gas content (e.g., \citealp{schmidt1959}; \citealp{kennicutt1998}). There is a power-law correlation between the surface densities of gas (\sgas) and SFR (\ssfr), with numerous studies attempting to constrain the power-law index via global quantities from galaxy samples and resolved observations of individual systems (see \citealp*{kennicutt2012} for a review). Spatially resolved data has uncovered that \ssfr \ is more strongly correlated with the surface density of molecular hydrogen (\shii) than atomic hydrogen (\shi) gas \citep*{wong2002}. More recent studies have begun to probe this relationship at sub-kpc scales, approaching the typical size of GMCs and the intrinsic physical scale of star formation, for larger samples of galaxies (e.g., \citealp{bigiel2008}; \citealp{momose2013}). Strong observational evidence also shows a correlation between the stellar mass and the gas-phase metallicity (e.g., \citealp{lequeux1979}; \citealp{tremonti2004}). Recent efforts are attempting to observationally constrain the relationships between stellar mass, metal content and \hi \ gas content (e.g., \citealp{hughes2013}; \citealp{bothwell2013}).

Studying the dust content has proved a difficult endeavour mainly due to technological limitations. Dust absorbs up to 50\% of the optical and ultraviolet (UV) photons from stars. The dust can then be traced via the re-emission of this energy at mid-infrared (MIR), far-infrared (FIR) and submillimetre (submm) wavelengths. Previous space missions - necessary to avoid atmospheric absorption in these bands - such as IRAS, ISO, \textit{Spitzer} and AKARI (respectively: \citealp{neugebauer1984}; \citealp{kessler1996}; \citealp{werner2004}; \citealp{murakami2007}) typically operated at wavelengths  $\lambda \sim$~240~$\mu$m, potentially missing a great part of the spectral energy distribution (SED) associated with a cold dust component. Fortunately, the success of the \textit{Herschel} Space Observatory \citep{pilbratt2010} satellite has substantially improved this situation. With two of its instruments, the Photodetector Array Camera and Spectrometer \citep[PACS,][]{poglitsch2010} and the Spectral and Photometric Imaging REceiver \citep[SPIRE,][]{griffin2010}, \textit{Herschel} observations are capable of constraining the FIR SED, and thus detect emission at $\lambda >$~250~$\mu$m originating from a cold dust component (e.g., \citealp{bendo2010b}), via imaging in six wavebands from 70 to 500~$\mu$m with significantly higher sensitivity and angular resolution than previous FIR/submm experiments.

Recent observational studies have used \textit{Herschel} data to investigate the correlations between the dust component and other galaxy properties (see e.g., \citealp{smith2012a}; \citealp{bourne2013}; \citealp{remyruyer2013}). For example, scaling relations have been found between the total masses of dust, H$_{2}$ and \hi \ in late-type Virgo cluster galaxies \citep{corbelli2012}, between the dust-to-stellar mass ratio and $NUV-r$ colour (\citealp{cortese2012}), and between the dust temperature and SFR per unit dust mass \citep{clemens2013}. A growing number of studies are using the unprecedented angular resolution of the \textit{Herschel} observations to perform pixel-by-pixel fitting of the FIR/submm SED and map the main dust properties of nearby galaxies (e.g., \citealp{smith2010}; \citealp{galametz2012}; \citealp{boquien2013}; \citealp{draine2013}; \citealp{tabatabaei2013}). Observations obtained by the \textit{Herschel} Exploitation of Local Galaxy Andromeda \citep{fritz2012} enabled highly-detailed maps of the dust properties of M31 to be produced via such pixel-by-pixel SED fitting \citep{smith2012}, and allow for an in-depth study of Andromeda's ISM (\citealp{ford2013}; \citealp[][in prep.]{viaene2013}; see also \citealp{draine2013}). Similarly, \citet{foyle2012} and \citet{mentuchcooper2012} have examined the gas and dust using \textit{Herschel} observations to carry out a pixel-by-pixel analysis of the inner regions of M83 and M51, respectively. Neither galaxy shows strong variations in the dust-to-gas mass ratio, except for a small gradient in M51 \citep{mentuchcooper2012}, whereas many other galaxy studies have found strong gas-to-dust ratio gradients (e.g., \citealp{munozmateos2009}; \citealp{bendo2010}; \citealp{magrini2011}). Although much effort has been made to quantify the radial dust distribution in these face-on spiral galaxies, we also require information on the vertical $|z|$ structure for a complete picture of the three dimensional properties of dust and gas in galaxies.  

Edge-on spiral galaxies with inclinations at or near 90$^{\circ}$ offer the best opportunity to study the radial and vertical structure of the various galaxy components. The stellar component has been shown to be sensitive to the merging history of a galaxy, whereas the ISM may probe gravitational instability (see e.g., \citealp{dalcanton2004}; \citealp*{yoachim2006}). Dust lanes feature prominently in optical images of massive edge-on spirals, enabling the study of dust in both emission and absorption. Furthermore, the increased surface brightness arising from the line-of-sight projection allows for both the vertical and radial dust distribution to be studied in detail at greater distances from the centre. Edge-ons therefore provide an important contribution to our knowledge of the dust structure. Fortunately, we have one such prototypical example of an almost perfect edge-on spiral right in our neighbourhood: NGC~891. Located at a distance of $\sim$9.6 Mpc (e.g., \citealp{strickland2004}) with an inclination angle of $>$89$^{\circ}$  (e.g., \citealp{xilouris1998}), this bright, non-interacting SA(s)b galaxy \citep{RC21976} presents similarities to our own Milky Way in Hubble type, rotational velocity and optical luminosity (see Table~\ref{tab:basicproperties}). These properties make NGC~891 an ideal target for studying the stars, gas and dust in the disk, and so it has already been extensively observed at a range of wavelengths. 

The stellar content in the galaxy, traced in the optical wavelengths (e.g., \citealp*{vanderkruit1981}; \citealp{ibata2009}) and at 3.6~$\mu$m \citep{kamphuis2007}, are found to be fairly symmetric along the galactic plane. However, the \hi \ gas shows a slight north-south asymmetry in both disk and halo thickness \citep{swaters1997}, and extends up to at least 22 kpc above the galactic plane \citep{oosterloo2007}. In contrast, molecular hydrogen gas observed via CO emission is predominantly found in a thin disk around 400 pc thick \citep{scoville1993}, and also seen in the halo up to 1.4 kpc above the plane \citep{gb1992}. The H$\alpha$ emission, a tracer of ionized gas linked to star formation, is more prominent and extended in the $|z|$ direction on the northern side of the galaxy than the southern (\citealp{dettmar1990}; \citealp{rand1990}). Whether the north-south asymmetry arises due to higher SFR in the north \citep{rossa2004} or increased attenuation by dust in spiral arms along the line-of-sight through the disk (\citealp{kamphuis2007}; \citealp{schechtmanrook2012}) remains an open issue, which may be resolved by studying the cold dust in the galaxy. 

Evidence for cold dust in NGC~891 first came from IRAS observations at FIR wavelengths (\citealp{wainscoat1987}), and later at millimetre wavelengths using the IRAM 30~m telescope (\citealp{guelin1993}). The 1.3~mm dust emission was noted to correlate strongly with the H$_{2}$ gas traced by CO emission, but poorly with the \hi \ emission. Later observations at $\lambda\,$ 450 and 850~$\mu$m with the Submillimeter Common-User Bolometer Array (SCUBA) at the James Clerk Maxwell Telescope (JCMT) found a large amount of cold dust ($\sim 15$\,K) in the disk \citep{alton1998}. The submm emission was confirmed to correlate spatially with the molecular gas in the disk (\citealp{alton1998}; \citealp{israel1999}). The combination of the SCUBA observations, IRAS and ISO data, with maps from the French balloon-borne PRONAOS experiment \citep{serra2002} traced the complete FIR/submm SED of NGC~891 for the first time \citep{dupac2003a}, finding a dust temperature of $\sim$ 18 - 24 K and estimating a global gas-to-dust mass ratio of $\sim$ 240. \citet{popescu2004} present deep maps of dust emission at  170 and 200~$\mu$m obtained with the ISOPHOT instrument. However, the low resolution of these datasets, typically $\geq$ 1\arcmin , meant a pixel-by-pixel analysis of the FIR/submm SED to map the cold dust distribution was not previously feasible. 

New, high resolution ($\leq$ $36\farcs4$) \textit{Herschel} PACS/SPIRE observations of NGC~891 have been obtained as part of the \textit{Herschel} Guaranteed Time Key Project the Very Nearby Galaxies Survey (P.~I.:~C.~D.~Wilson), which aims to study the gas and dust in the ISM of a diverse sample of 13 nearby galaxies using \textit{Herschel}. These observations now make it possible to trace the FIR/submm SED in detail and map the resolved dust properties in the galaxy disk. In this paper, we investigate the connection between dust and gas in NGC~891 in order to test for asymmetries in the distributions of the dust properties and the gas-to-dust ratio. We use these new \textit{Herschel} PACS/SPIRE observations combined with a JCMT SCUBA 850~$\mu$m image to trace the FIR/submm SED. For the first time, we fit one-component modified blackbody models to the SEDs of each pixel to map the dust mass and temperature in the galaxy. HI 21 cm line emission and $^{12}$CO(J=1-0) emission maps, combined with a new JCMT HARP-B $^{12}$CO(J=3-2) map (P.~I.:~T.~Parkin), trace the atomic and molecular hydrogen gas, enabling the study of the gas-to-dust ratios of these components. Our paper is structured as follows. In the next section, we present the new and existing data used in our analysis. In Section 3, we present our detection of extraplanar dust. In Section \ref{sec:dustprops}, we describe our SED fitting methodology and the resulting integrated and resolved dust properties. Section \ref{sec:dustgas} describes the relationships we find between the dust and gas, and Sections \ref{sec:discussion} and \ref{sec:conclusions} present our discussion and conclusions, respectively.

\begin{table}
 \centering
 \begin{minipage}{\columnwidth}
  \caption{Basic properties of NGC 891}
  \label{tab:basicproperties}
  \begin{tabular}{lll}
  \hline
\hline
Property & value & reference\\
\hline
RA (J2000) & $2^\mathrm{h}\ 22^\mathrm{m}\ 33\fs0$ & NED\footnote{The NASA Extragalactic Database (NED) is available online at \\ \url{http://nedwww.ipac.caltech.edu/}} \\
Dec (J2000) & +$42\degr\ 20\arcmin\ 57\farcs2$ & NED \\
Hubble type & SA(s)b & \citet{RC21976} \\
distance & 9.6 Mpc & Strickland et al. (2004) \\
optical major axis & 13.5 arcmin & \citet{RC31991} \\
optical minor axis & 2.5 arcmin & \citet{RC31991} \\
major axis angle & $22^\circ$ & NED \\
inclination angle & $89^\circ$ & \citet{xilouris1998} \\
B-band luminosity & 7.8 $\times$ $10^{9}$ $\mathrm{L}_\odot$ & \citet{RC31991} \\
rotational velocity & 225 km s$^{-1}$ & \citet{vanderkruit1984} \\
total mass & 1.4 $\times$ $10^{11}$ $\mathrm{M}_\odot$ & \citet{oosterloo2007} \\
SFR & 3.8 $\mathrm{M}_\odot$ yr$^{-1}$ & \citet{popescu2004}\\    
\hline
\end{tabular}
\end{minipage}
\end{table}

\newpage

\section{Observations}\label{sec:data}

We have assembled together a multi-wavelength dataset capable of tracing the key ISM components in NGC~891 (see Fig. \ref{fig:maps} and Table \ref{tab:obsproperties}). Our new VNGS \textit{Herschel} PACS photometric observations and a new JCMT {HARP-B} \mbox{$^{12}$CO(J=3-2)} emission map are presented here for the first time.

\subsection{Mid-infrared}


{\it Spitzer} Infrared Array Camery (IRAC; \citealp{fazio2004}) 3.6~$\mu$m data were obtained in astronomical observation requests 3631872, 3632128, 3632384, and 3633152.  Individual corrected basic calibration data frames were processed with version 18.25.0 of the IRAC pipeline and remosaicked using the standard IRAC pipeline within the MOsaicker and Point source EXtractor \citep{makovoz2005}.  The final images have pixel scales of $0\farcs75$, and the images are oriented with north up and east to the left.  The PSFs have FWHM of $1\farcs7$ and calibration uncertainties of 3\% (IRAC Instrument and Instrument Support Teams, 2013, IRAC Instrument Handbook, Version 2.0.3, JPL, Pasadena).


The Wide-field Infrared Survey Explorer (WISE; \citealp{wright2010}) imaged NGC~891 as part of an all-sky survey in four bands: 3.4, 4.6, 12, and 22~$\mu$m. For the purposes of our analysis, we only use the latter two MIR bands. The angular resolution is $6\farcs5$ and $12\farcs0$ at 12 and 22~$\mu$m, and the astrometric precision for high signal-to-noise sources is better than $0\farcs15$. Although deconvolution techniques may enhance the WISE image resolution, the mosaics with standard beams are sufficient for our analysis.


Multiband Imaging Photometer for {\it Spitzer} (MIPS; \citealp{rieke2004}) 24~$\mu$m data were reprocessed by \citet{bendo2012b} using the MIPS Data Analysis Tools \citep{gordon2005} along with additional processing steps.  The final mosaic has a pixel scale of $1\farcs5$, the PSF FWHM is $6\arcsec$, and the calibration uncertainty is 4\% \citep{engelbracht2007}.

\subsection{Far-infrared}


VNGS {\it Herschel} PACS observations at  70 and 160~$\mu$m comprised four cross-scans each, taken at a 20$\arcsec$ s$^{-1}$ scan rate. We also include an additional PACS observation at 100~$\mu$m, taken as part of The \textit{Herschel} EDGE-on galaxy Survey (HEDGES; P.I.:~E.~Murphy), which consists of only two cross-scans at a 20$\arcsec$~s$^{-1}$ scan speed (Obs IDs: 1342261791, 1342261793) but with a smaller separation between individual scan legs. The \textit{Herschel} Interactive Processing Environment \citep[HIPE, v.11.0.1;][]{ott2010} with PACS Calibration (v.32) was first used to bring the raw Level-0 data to Level-1, which means flagging of pixels, flat-field correction, flux conversion into units of Janskys, and assigning sky coordinates to each detector array pixel. Scanamorphos \citep[v.22.0,][]{roussel2012} was then used to remove the 1/$f$ noise, drifts and glitches by exploiting the redundancy of the observations in each sky pixel to estimate the optimal baseline correction for each detector pixel. This process produced the final images. The maps have pixel sizes of $1\farcs4$, $2\farcs0$ and $2\farcs85$ for the 70, 100 and 160~$\mu$m maps respectively. These pixel sizes correspond to one quarter of the point spread function (PSF) full width at half maximum (FWHM) for the scan speed used  
\clearpage
   
\begin{figure*}[!H]
\begin{center}
\vspace{1cm}
\includegraphics[width=\textwidth]{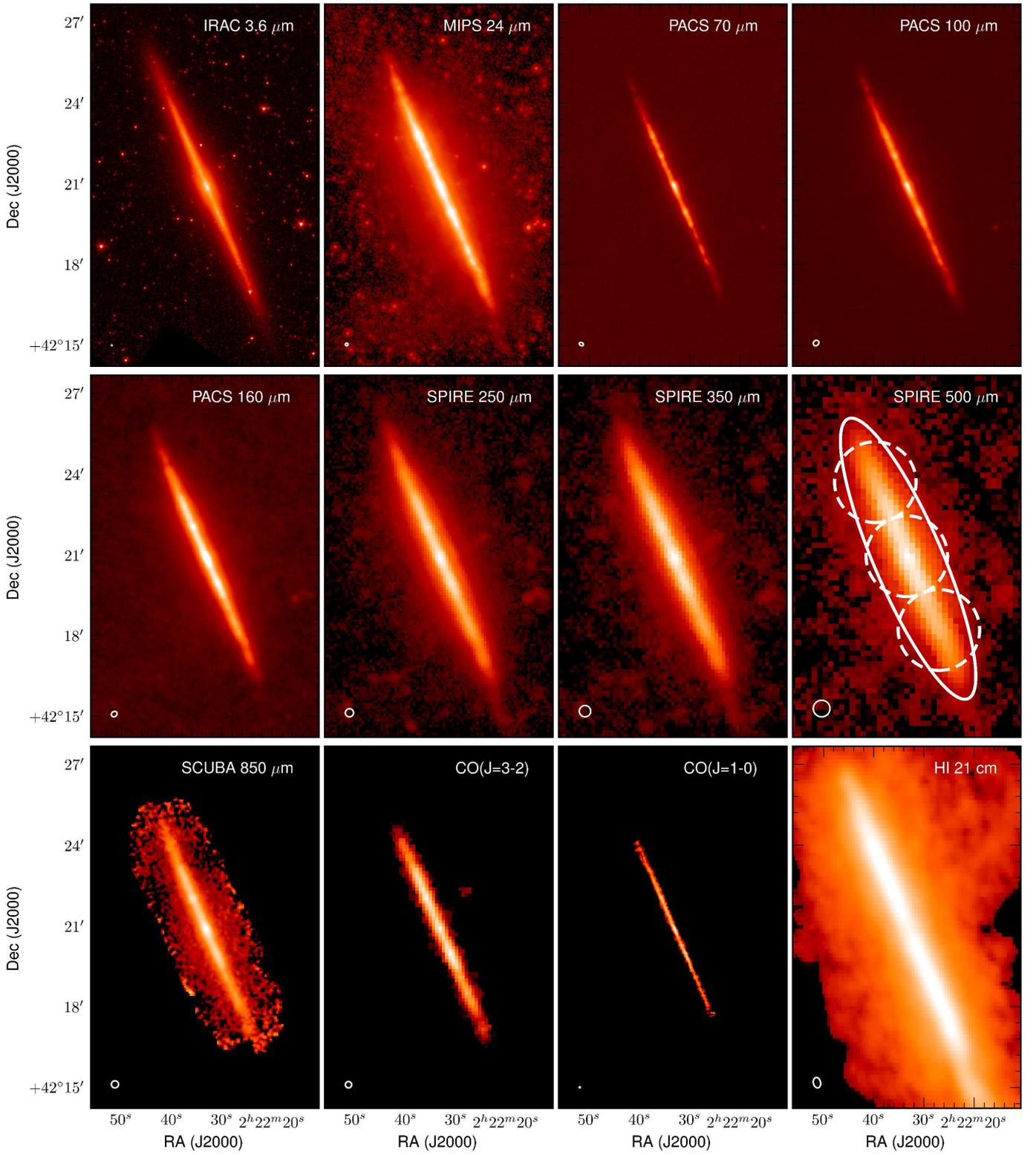}
\end{center}
\caption[Observations]{Images of NGC~891 at various wavelengths, centred on the galaxy's coordinates from NED: $\alpha = 2^\mathrm{h}\ 22^\mathrm{m}\ 33\fs0$, 
$\delta = 42\degr\ 20\arcmin\ 57\farcs2$ (J2000.0). Each image is presented in its native resolution and pixel size, with the open symbols in the bottom left of each panel indicating the beam size at each wavelength. The image characteristics are described in detail in the text (see Section \ref{sec:data}). The solid white ellipse denotes the aperture for integrated photometry and the dashed white circles represent the apertures to reconstruct the analysis of \citet{dupac2003b}, as described in Section \ref{sec:intphot}. North is up and east is to the left.}\label{fig:maps}
\end{figure*}

\clearpage
\noindent
in our observations (PACS Observers' Manual, 2013). We apply a colour correction factor, taken from the ICC “PACS Photometer Passbands and Colour Correction Factors” document\footnote{See \url{http://herschel.esac.esa.int/twiki/pub/Public/}\newline \url{PacsCalibrationWeb/cc_report_v1.pdf}}, adopting values relative to a modiﬁed black body with temperature of 20 K and emissivity index $\beta =$ 2, which are both appropriate for the SED of NGC 891 (see Section~\ref{sec:dustprops}).

\subsection{Submillimetre}


VNGS {\it Herschel} SPIRE photometric observations were taken at 250, 350, and 500~$\mu$m. The galaxy was observed in large map mode, covering an area of 20$\arcmin$x20$\arcmin$ with two cross-scans at a 30$\arcsec$ s$^{-1}$ scan rate. 
The SPIRE data were processed up to Level-1 using a script based on the standard pipeline, and was run in HIPE continuous integration build number 11.0.1200. There are two differences to the standard pipeline: firstly, we use the latest calibration due for inclusion in HIPE 12.0 \citep{bendo2013}, and secondly we do not apply the default temperature drift correction and de-striping. Instead we use a custom method (BriGAdE; M. W. L. Smith in prep.) to remove the temperature drift and bring all the bolometers to the same level. SPIRE instrument characteristics, e.g. PSF, beam size, calibration uncertainty, etc., are taken from the SPIRE Observers' Manual (v.2.4, 2011).  Artefacts introduced due to extended bright sources were avoided by masking the galaxy prior to removing the median baselines from each bolometer timeline. The HIPE mapmaking procedure was then used to produce the maps. The main beam size has a FWHM of $18\farcs1$, $24\farcs9$ and $36\farcs4$ at 250, 350 and 500~$\mu$m maps, respectively, and the resulting images have pixel sizes of 6, 8 and 12$\arcsec$. We note that in addition to the use of more up-to-date versions of the pipeline and calibration tree compared to the maps previously presented in \citet*{bianchi2011}, we adopt new beam areas of 446, 794 and 1679 arcsec$^{2}$ at 250, 350 and 500~$\mu$m, respectively\footnote{See \url{ftp://ftp.sciops.esa.int/pub/hsc-calibration/}\newline \url{SPIRE/PHOT/Beams/beam_release_note_v1-1.pdf}}. We also apply a calibration factor to account for extended sources and updated colour corrections based on an index of $\alpha_{s} =$~4 (SPIRE Data Reduction Guide, v.2.3, 2013), which is appropriate for $\beta =$~1.8 (see Section \ref{sec:sedfitting}).


We use the JCMT SCUBA observations at 850~$\mu$m presented in \cite{alton1998}. The dedicated SCUBA software package SURF \citep{jenness1998}, was used to clean, flat-field, and calibrate the image according to atmospheric attenuation. The measured half-power beam width of the central beam is $15\farcs7$ at 850~$\mu$m, with sidelobes estimated as 20\% (from the same observing configuration, see \citealp{bianchi1998}). \citeauthor{alton1998} estimate the error in the calibration to be less than 15\% with a 1-$\sigma$ noise level of 3.5 mJy/$15\farcs7$ beam at 850~$\mu$m. The original data from \citet{alton1998} has since been recalibrated and presented in \cite{haas2002} and \cite{whaley2009}. Residual noise seen at the SCUBA field edges is avoided in our analysis (see Section \ref{sec:intphot}). We apply a conservative correction of 5\% to the flux density to account for contamination due to the \mbox{$^{12}$CO(J=3-2)} line (see e.g., \citealp{israel1999}).

\subsection{CO and \hi \ line data}

We obtained new $^{12}$CO(J=3-2) observations, which were carried out on 19 December 2009 and 25 January 2010 using the HARP-B receiver mounted on the JCMT, as part of project M09BC05 (P.~I.: T. Parkin). We used quarter-step spacing and the basket weave pointing technique (see the Appendix of \citealp{warren2010}) to produce a raster map of the galaxy $3.25\arcmin \times 8.75\arcmin$ in size, at a resolution of 14$\arcsec$. The total integration time required for this observation was 2180~s. The sky background was quantified by position-switch chopping of the secondary mirror with a chop throw of 60$\arcsec$. We used the Auto-Correlation Spectrometer Imaging System (ACSIS) for the backend receiver, set to a bandwidth of 1000~MHz with 2048 channels. This configuration produces a spectral resolution of 0.43~km~s$^{-1}$. We processed the raw data with Starlink\footnote{The \textsc{Starlink} package is available for download at http://starlink.jach.hawaii.edu.} \citep{currie2008}. These data were flagged for bad pixels, combined into a data cube with a pixel-size of $7\farcs2761$, and then baseline subtracted by fitting a third order polynomial to the line-free regions of the spectrum. Finally, the cube was rebinned in the spectral dimension to a resolution of 20~km~s$^{-1}$. The average root mean-square (rms) noise in the spectrum then produced a signal-to-noise cube. The integrated intensity map was created by running the Starlink routine `find-clump' on the signal-to-noise cube to identify regions of emission with a S/N $>$ 3. This cube was collapsed along the spectral axis to produce a two-dimensional signal-to-noise integrated intensity map, which was subsequently multiplied by the noise to produce the final integrated intensity map. The integrated intensity measurement uncertainty is a function of the channel width, the rms noise, and the number of channels in the spectral line and baseline. The calibration uncertainty is estimated to be 10\% based on a comparison of spectra taken of calibrators during observations and their standard spectra. For more details on the data reduction process and uncertainty measurements, see \citet{warren2010} and \citet{parkin2012}.


A high resolution map of the $^{12}$CO(J=1-0) emission is taken from \cite{scoville1993}. The map was made using the Owens Valley Millimeter Array (OVMA) to observe nine fields along the semimajor axis, covering $\pm 3\farcm6$ from the centre. The synthesized beam from the different telescope configurations ranged from $2\farcs2$ to $2\farcs6$ with a mean of $2\farcs3$. The spectral resolution is based on 32 channel filter banks with 5 MHz resolution and the continuum calibration was performed with a broad-band channel with an effective bandwidth of 375 MHz. Observations of 3C 84 and Uranus provided the phase and absolute flux calibrations. The nine fields were cleaned and mosaicked. The final map has a rms noise of 20 mJy beam$^{-1}$. 


Finally, neutral hydrogen 21 cm line emission data necessary for quantifying the \hi~content was available from \cite{oosterloo2007}, taken as pathfinder observations for the Hydrogen Accretion in LOcal GAlaxieS survey (HALOGAS; \citealp{heald2010}, \citeyear{heald2011}) with the Westerbork Synthesis Radio Telescope (WSRT). Standard array configurations for the WSRT were used to complete 20 $\times$ 12 hr exposures, with a total bandwidth of 10 MHz using 1024 channels. The data was reduced using the MIRIAD package \citep{sault1995}. We use their intermediate resolution data, which has a beam FWHM of 30$\arcsec$ \citep[see Fig 1. in][]{oosterloo2007}. The map is the deepest \hi \ observation available for NGC~891. 

\begin{table}
 \centering
 \begin{minipage}{\columnwidth}
  \caption{Summary of observations of NGC 891.}
  \label{tab:obsproperties}
  \begin{tabular}{l c c  c c c}
\hline 
\hline
Instrument & $\lambda$    &  pixel & PSF  & calibration & ref.\footnote{References are (1)~IRAC Instrument Handbook, 2013; (12)~\citealp{jarrett2013}; (3)~\citealp{bendo2012b}; (4)~PACS Observers' Manual, 2011; (5)~SPIRE Observers' Manual, 2011; (6)~\citealp{alton1998}; (7)~\citealp{haas2002}; (8)~this work; (9)~\citealp{scoville1993}; (10)~\citealp{oosterloo2007}.} \\
           &              &  size &  FWHM  & uncertainty &           \\      
\hline
 IRAC      &  3.6~$\mu$m  &  $0\farcs75$     &  $1\farcs7$          & 3\%  & (1) \\  
 WISE      &  12~$\mu$m   &   $1\farcs4$     &  $6\farcs5$          & 5\% & (2) \\ 
 WISE      &  22~$\mu$m   &   $1\farcs4$     &  $12\farcs0$         & 6\% & (2) \\ 
 MIPS      &  24~$\mu$m   &   $1\farcs5$     &  $6\farcs0$          & 4\%  & (3) \\ 
 PACS      &  70~$\mu$m   &    $1\farcs4$    &  $5\farcs6$          & 5\%  & (4) \\ 
 PACS      &  100~$\mu$m  &    $2\farcs0$    &  $8\farcs0$          & 5\%  & (4) \\ 
 PACS      &  160~$\mu$m  &    $2\farcs85$   &  $11\farcs4$         & 5\%  & (4) \\ 
 SPIRE     &  250~$\mu$m  &   $6\farcs0$     &  $18\farcs1$         & 7\%  & (5) \\ 
 SPIRE     &  350~$\mu$m  &   $8\farcs0$     &  $24\farcs9$         & 7\%  & (5) \\ 
 SPIRE     &  500~$\mu$m  &   $12\farcs0$    &  $36\farcs4$         & 7\%  & (5) \\ 
 SCUBA     &  850~$\mu$m  &   $3\farcs0$     &  $15\farcs7$         & 15-25\%  & (6)-(7)\\ 
 HARP-B    &  869~$\mu$m  &   $3\farcs0$     &  $14\farcs0$         & 10\% & (8) \\
 OVMA      &  2.6~mm      &   $0\farcs7$     &  $2\farcs3$          & 5\%  & (9)\\ 
 WSRT      &  21~cm       &   $9\farcs0$     &  $30\farcs0$         & 3\%  & (10)\\ 
\hline
\end{tabular}
\end{minipage}
\end{table}

\section{FIR/submm morphology}\label{sec:submmmorph}

Before proceeding with our main analysis, we first take a brief diversion to exploit one of the main advantages of studying an edge-on galaxy like NGC 891 - the ability to study the vertical and radial distribution of the FIR/submm emission.

\subsection{Detection of extraplanar dust}\label{sec:extraplanar}

The presence of a spatially-resolved vertical profile in the FIR/submm emission of an edge-on galaxy may be related to dust ejected from the galactic plane, i.e. extraplanar dust. Determining the prevalence of extraplanar dust and the relationship to other galaxy properties may shed light on the mechanisms responsible for transporting ISM material from the galactic plane. Deep optical to NIR imaging has revealed significant amounts of extraplanar dust (e.g., \citealp{howk1999}; \citealp{thompson2004}) as well as PAHs and small dust grains (e.g., \citealp{whaley2009}) in several edge-on spirals, including NGC 891. Most recently, extraplanar dust was detected at the 5-$\sigma$ level in the small edge-on spiral NGC 4244 \citep{holwerda2012a} and one (NGC 4013) of the seven edge-on spirals comprising the \textit{HERschel} Observations of Edge-on Spirals \citep[\textit{HER}OES, ][]{verstappen2013}. 

\begin{figure}
\begin{center}
\includegraphics[width=\columnwidth]{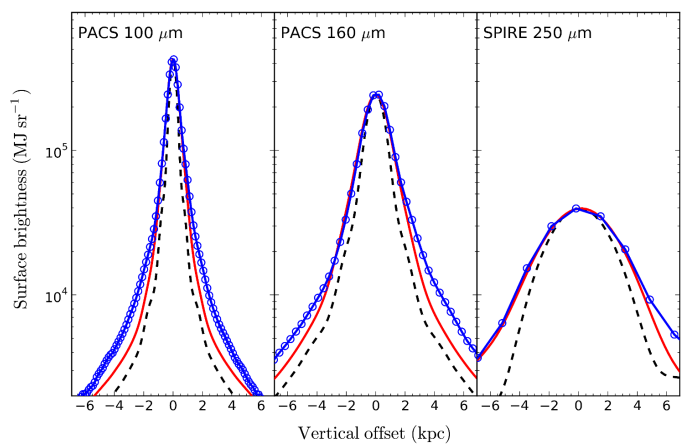}
\end{center}
\vspace{-0.3cm}
\caption[FIR vertical profiles]{Vertical profiles at 100, 160 and 250~$\mu$m, adopting a position angle of 22.9$^{\circ}$, are plotted as blue circles connected by a blue line. We fit with an exponential profile (solid red line), which has been convolved with the \textit{Herschel} beam at the corresponding wavelength (black dashed line).}\label{fig:vertiprofs}
\end{figure}

We extracted vertical profiles of NGC 891 from the PACS 100, 160 and SPIRE 250 $\mu$m images. We selected an area surrounding the galaxy and summed all the pixel values along the major axis to generate the profiles. Following \citet{verstappen2013}, we model these profiles with an exponential function appropriate for an exactly edge-on, double-exponential disc, given by
\begin{equation}\label{eqn:vertiprof}
\Sigma_{\mathrm{ver}}(z)=\frac{1}{2 h_z}\exp\left(-\frac{|z|}{h_z}\right)
\end{equation}
where $h_z$ is the dust scaleheight. We first convolve the vertical profile model with the \textit{Herschel} beams at each corresponding wavelength, using the circularised PSF images\footnoteremember{aniano}{PSFs, convolution kernels and the IDL task \textsc{convolve\_image.pro} from Aniano et al. are available from \url{http://www.astro.princeton.edu/~ganiano/Kernels.html}.} described in \citet{aniano2011}. In order to obtain one-dimensional beams, the two-dimensional PSFs are averaged along one direction in the same manner as we obtain the vertical profiles. The optimal value of $h_z$ that reconciles the observed and model profiles was found by using a $\chi^2$ minimisation technique. Uncertainties on the value of $h_z$ were also derived from the $\chi^2$ probability distribution. We estimate the uncertainty in the profiles due to the galaxy's position angle ($PA$), which varies in the literature from 22 to 23$^{\circ}$, by comparing the resultant $h_z$ values and profiles when systematically varying the assumed position angle from 21 to 24$^{\circ}$ in 0.1$^{\circ}$ increments. For position angles moving out of the range of 22.6 $<\ PA\ <$ 23.1$^{\circ}$, the profiles demonstrate increased broadening and an expected increase in the corresponding $h_z$ value. The profile widths and scaleheights tend to reach minimum values within the 22.6 $<\ PA\ <$ 23.1$^{\circ}$ range. Combining these results, the recent result from PA fitting performed by \citet{bianchi2011}, and a visual inspection of the data, we decide to adopt $PA\ =$ 22.9$^{\circ}$ for our analysis.

The resulting vertical profiles are shown in Fig.~\ref{fig:vertiprofs}. We observe a slight asymmetry in the profiles towards the NW side of the disk (represented by positive vertical offsets in Fig.~\ref{fig:vertiprofs}), which are also seen in the submm profiles obtained by \citet[][see their Fig.~4]{bianchi2011}. For NGC 891, we derive FIR scaleheights of 0.24$^{+0.05}_{-0.04}$, 0.43$^{+0.06}_{-0.05}$ and 1.40$^{+0.20}_{-0.24}$~kpc at 100, 160 and 250~$\mu$m, respectively. The scaleheight at 100~$\mu$m is consistent with scaleheights found in previous studies using radiative transfer modelling of optical and NIR images. For example, \citet{xilouris1998} found a mean dust scaleheight of 0.26~kpc with a range of 0.22 to 0.32~kpc from models based on B to K band images. Similar results were recently found in the modelling by \citet{schechtmanrook2012}. In such a comparison of scaleheights derived from different methods, we should not forget that (i) the FIR profiles have a smaller resolution (e.g. $\sim7\arcsec$ at 100~$\mu$m) and are much more dominated by the PSF FWHM than compared to optical images used for the radiative transfer modelling ($\sim1\arcsec$), and (ii) our implicit assumption of a perfectly edge-on disk likely differs from the inclination adopted by the radiative transfer models. However, NGC 891's near perfect edge-on inclination (89.7$^{\circ}$, e.g., \citealp{xilouris1998}) means the latter issue is likely negligible. 

Given that the profiles are not dominated by the telescope beam, as evident in Fig.~{\ref{fig:vertiprofs}}, and since the deconvolved scaleheight value we derive from the profile fitting is \textit{not} consistent with zero at the 5-$\sigma$ level, we conclude that our vertical FIR profiles are spatially resolved. Such a conclusion confirms previous findings of extraplanar dust in NGC 891 inferred from optical-NIR imaging (see e.g., \citealp*{howk1999}). However, whether this material has been expelled from the galactic plane or infalling from the intergalactic medium (e.g., \citealp{howk2009}) is still not understood.

\subsection{Break in PACS radial profiles}

\begin{figure}
\begin{center}
\includegraphics[width=\columnwidth]{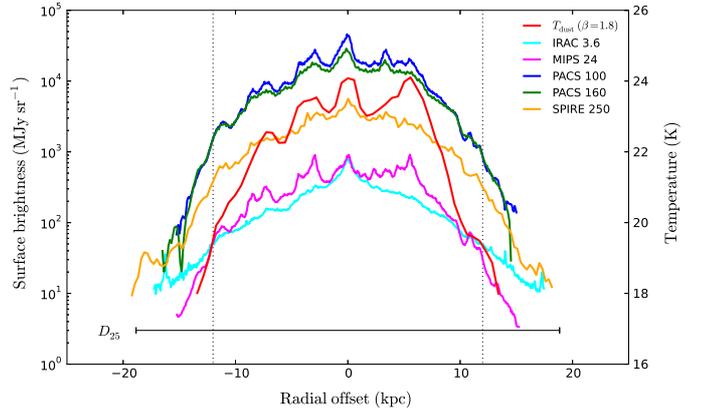}
\end{center}
\vspace{-0.3cm}
\caption[FIR radial profiles]{Radial profiles at 100, 160 and 250~$\mu$m are compared with the IRAC 3.6~$\mu$m and MIPS 24~$\mu$m emission. The solid black line represents the optical diameter adopting $D_{25}\ =$~37.7~kpc and the two dashed black lines demark the break radius found at $\pm$12 kpc from \citet*{bianchi2011}. The alternative y-axis scale refers only to the dust temperature profile obtained from modelling the SED (see Section~\ref{sec:sedfitting}), represented by the solid red line.}\label{fig:radialprofs}
\end{figure}

A break in the radial profiles in the SPIRE bands, found by \citet*{bianchi2011}, was shown to occur in all bands at $\sim$ 12 kpc from the centre, with the profiles sharply declining at larger radii with a reported radial scalelength of about 1 kpc. A similar break was also detected in the optical images of \citet{xilouris1999}. We now investigate whether we observe the same behaviour in the radial profiles of the PACS bands. We produce horizontal profiles of the FIR/sub-mm emission from the PACS and SPIRE data via a similar method to that described for extracting the vertical profiles. The flux was summed along pixels parallel to the minor axis and divided by the number of pixels used for each strip, thus yielding an average surface brightness profile. The resulting profiles are presented in Fig.~\ref{fig:radialprofs}. The SPIRE 250 $\mu$m profile matches that found by \citet*{bianchi2011}. Our profiles from the PACS bands broadly display similar features as the SPIRE profiles: a central peak with two clear secondary peaks within radial distances of approximately $\pm$5 kpc, a sharp decrease in surface brightness between $\pm$11 and 13 kpc, and a flattening of the profile at radial distances greater than $\pm$13 kpc. Thus, we also observe the same break in the PACS bands as the analysis of the SPIRE profiles performed by \citet*{bianchi2011}.

The most interesting observation to note is the correlation between the profiles of the IRAC 3.6~$\mu$m emission and the MIPS 24~$\mu$m emission, which respectively trace the total stellar population and star formation. Within the break radius at $\pm$12 kpc (dashed line in Fig.~\ref{fig:radialprofs}) the FIR profiles appear to correlate best with the 24~$\mu$m emission. At radial distances outside the break radius, the FIR profiles on the SW edge of the disc (i.e. negative radial offsets) appear to flatten out to reflect the 3.6~$\mu$m emission, whereas the profiles on the NE edge (i.e. positive radial offsets) are only traceable to $\sim$ 85 \% of the optical size. This appears to hint that the dust properties change across the break radius, possibly as the dominant source of the dust heating may changes from star-forming regions to the stellar population. We return to the topic of the source of dust heating in more detail in the following section.

\section{Dust properties of NGC~891}\label{sec:dustprops}

We now present our methodology to determine the global and spatially-resolved dust properties of NGC~891.

\subsection{Image preparation}\label{sec:imagemani}

In our analysis, we fit simple models to both the integrated SED of the galaxy and to the individual SEDs of each pixel, so that we may derive the global dust properties and map their spatial distributions. However, the various images described above were taken using several different instruments and detectors, with differing PSFs of varying shapes and widths. To properly compare the data to each other and perform a pixel-by-pixel comparison, the physical scale of each image should be consistent. We therefore require convolution kernels that will transform all the images into a common PSF, so we can generate image cubes in which each pixel of each image corresponds to the same sky region. Fortunately, common-resolution convolution kernels have already been made publicly available\footnoterecall{aniano} for the instrumental PSFs of several space- and ground-based telescopes, as well as general purpose Gaussian PSFs \citep{aniano2011}. We use the dedicated kernels for PACS, SPIRE and WISE images, which were generated with the latest available PSF characterization for each instrument, and use the appropriate Gaussian kernels for the remaining images. 

All images were first converted into surface brightness units (i.e. MJy sr$^{-1}$) and convolved to the resolution of the 500~$\mu$m image, since this band has a PSF with the largest FWHM ($36\farcs4$), using the IDL task \textsc{convolve\_image.pro} \citep{aniano2011}. The images were regridded to the pixel size of the 500~$\mu$m map using the {\sc Montage} software package. We note that since the pixel size ($12\farcs$) is smaller than the 500~$\mu$m beam size ($36\farcs4$), adjacent pixels are not independent. Errors on each pixel were calculated by summing the flux calibration uncertainty, instrumental noise and sky background measurement in quadrature. For the pixels covering the galaxy, the flux errors are dominated by the calibration uncertainty. We use calibration uncertainties of 5\% for PACS (PACS Observers' Manual, 2011), 7\% for SPIRE (SPIRE Observers' Manual, 2011) and 25\% for SCUBA \citep{haas2002}. The sky background error was measured using the standard deviation of the flux from ten circular background apertures, which were carefully placed around the galaxy at a suitable distance to avoid diffuse galaxy emission. For each image, the mean sky flux estimated from these background apertures was subtracted.

\subsection{Source of dust heating}\label{sec:dustheatinganalysis}

Before fitting the SED data for NGC~891, we assessed how the dust was related to its heating sources. Several authors (e.g., \citealp{bendo2010b}, \citeyear{bendo2012a}; \citealp{boquien2011}) have compared dust colour variations to emission from either star forming regions or the total stellar populations and have demonstrated emission seen at $\geq 250$~$\mu$m by {\it Herschel} may originate from a different thermal component than the emission seen at $\leq 160$~$\mu$m. We perform a similar analysis on the NGC~891 data by comparing the 100/160, 160/250, 250/350, and 350/500~$\mu$m flux ratios to 3.6~$\mu$m emission (a tracer of the total stellar population that is relatively unaffected by dust extinction) and 24~$\mu$m emission (a tracer of star formation that is also unaffected by dust extinction e.g., \citealp{calzetti2007}). The galaxy is viewed edge-on, which may cause some issues with properly relating dust emission to heating sources, as emission from both star forming regions and diffuse emissions is integrated along the line of sight. On the other hand, the distribution of starlight and star forming regions is distinctly different in this galaxy. Emission from star formation peaks in the centre and in a couple of star forming regions in the disc, while 3.6~$\mu$m emission is relatively flat within the centre (because of the presence of the galaxy's bulge) and then smoothly decreases with radius (see Fig.~\ref{fig:maps}). Because of this, associating colour variations with dust heating sources is relatively straightforward.

\begin{figure}
\begin{center}
\includegraphics[width=0.99\columnwidth]{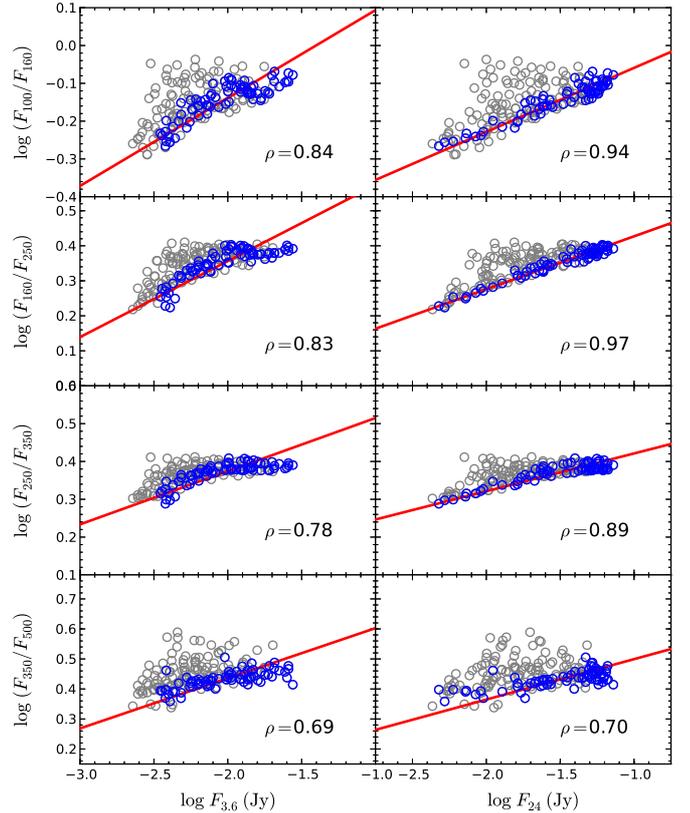}
\end{center}
\vspace{-0.3cm}
\caption[FIR/submm colours versus 3.6 and 24 micron emission]{The 100/160, 160/250, 250/350 and 350/500~$\mu$m flux ratios compared with the 3.6 and 24~$\mu$m emission. Pixels along the major axis shown in blue and off-axis pixels shown in grey. The best-fit linear relation (red line) and the Pearson correlation coefficient are based only on the on-axis pixels.}\label{fig:heating}
\end{figure}

The results are shown in Fig.~\ref{fig:heating}, with data along the major axis shown in blue and data off-axis shown in grey.  The off-axis pixels tend to exhibit scatter, potentially because of either low signal-to-noise in the data or issues with matching the extended structure in the PSFs following the convolution process described above, so we focus on the on-axis pixels. The 100/160, 160/250, and 250/350~$\mu$m ratios all correlate better with the 24~$\mu$m emission than with the 3.6~$\mu$m emission. In particular, the relations between the 100/160, 160/250, and 250/350~$\mu$m ratios and the 3.6~$\mu$m emission flatten out at high surface brightnesses, which indicates that the colours are independent of the galaxy's stellar surface brightness. This suggests that the 100-250~$\mu$m emission originate from a component of dust heated by star forming regions. The results from the 350/500~$\mu$m ratios are more ambiguous. It is possible that the dust seen in this wavelength range is heated by a combination of light from star forming regions and diffuse light from the total stellar population. However, it is also possible that the dust emitting at $\geq 350$~$\mu$m is still primarily heated by star forming regions and that the colour variations are difficult to detect because the emission originates from the Rayleigh-Jeans side of the SED.

These results suggest that, following integration along the line-of-sight, emission from dust in NGC~891 at $>250$~$\mu$m heated by the diffuse interstellar radiation field is either weak or absent in the middle of the plane. This would contradict the results from other spiral galaxies (e.g., \citealp{bendo2010b}, \citeyear{bendo2012a}; \citealp{boquien2011}; \citealp{groves2012}; \citealp{smith2012}), which all found significant amounts of emission from dust heated by the total stellar population. However, a scenario where dust heating in NGC 891 is dominated by star-forming regions rather than the total stellar population could explain why the models of \citet*{bianchi2011} required a stellar disk with radial scalelength of 5.7 kpc, more similar to that of young stars (traced via B-band imaging), instead of a 4 kpc radial scalelength derived from the old stellar population (emitting in NIR bands), to successfully reproduce the major axis profiles in the SPIRE bands. Furthermore, the results for NGC~891 would be consistent with those found for the dwarf irregular galaxy NGC 6822 \citep{galametz2010}. Bendo et al. (in preparation) are performing a follow-up analysis on dust heating over a broader range of spiral galaxies and will discuss additional galaxies where dust may be heated primarily by star forming regions.

\subsection{SED fitting}\label{sec:sedfitting}

Based on the above analysis, it is appropriate to follow previous works (e.g., \citealp{smith2010}, \citeyear{smith2012}; \citealp{verstappen2013}) in fitting the SEDs with a one component modified blackbody model given by
\begin{eqnarray}
F_{\nu} = \frac{M_{\mathrm{dust}}\ \kappa_{\nu}}{D^{2}} B_{\nu}(T_{\mathrm{dust}}) \mathrm{,}
\end{eqnarray}
\noindent
which is a modified version of the equation originally presented by \citet{hildebrand1983}. In this equation, $M_{\mathrm{dust}}$ is the dust mass, $B_{\nu}$ is the Planck function, $T_{\mathrm{dust}}$ is the dust temperature, $\kappa_{\nu}$ is the dust emissivity and $D$ is the distance to the galaxy (adopting 9.6~Mpc, \citealp{strickland2004}). We perform this fit on all data from 100 to 500~$\mu$m and, following an approach suggested by several authors (e.g., \citealp{bendo2010b}; \citealp{smith2010}, \citeyear{smith2012}), treat the 70~$\mu$m emission, which may originate from small grains with stochastic dust heating, as an upper limit to the fit. Although some of the individual surface brightness ratios for the off-axis regions shown in Fig.~\ref{fig:heating} exhibited some scatter, the results from the SED fitting across the 100-500~$\mu$m should mitigate the effects that noise or PSF mismatch issues in any individual band will have on characterising the overall shape of the SED. 

We assume a power-law dust emissivity in the FIR/submm wavelength range, with $\kappa_{\nu} \propto \nu^{\beta}$. Unfortunately, the absolute value of the dust emissivity and its frequency dependence are still uncertain (e.g., Dupac et al. 2003; Gordon et al. 2010; Planck Collaboration et al. 2011). A variation in the value of $\kappa_{\nu}$ translates the model SED flux densities, systematically increasing or decreasing the dust property estimates without affecting any correlations that we may find. Following previous works, we adopt $\kappa_{\nu}$ = 0.192 m$^{2}$ kg$^{-1}$ at 350~$\mu$m \citep{draine2007}, but note that this may vary as a function of $\beta$ (see e.g., \citealp{bianchi2013}, and references therein). The choice of value for $\beta$ is also not so straightforward, because this can potentially create both systematic offsets in the dust property estimates and introduce secondary effects in correlations between the dust mass/temperature estimates, and other quantities we wish to study, such as e.g. the various components of the total gas content.  Such effects can lead to an unreliable analysis. There is still much debate about whether the value of $\beta$ should have a fixed value, or vary as a free parameter, as the dust emissivity spectral index may provide information about the physical properties of the radiating dust grains (see e.g., \citealp{smith2012}). In our analysis, we primarily chose to allow $M_{\mathrm{dust}}$, $T_{\mathrm{dust}}$ and $\beta$ to vary as free parameters, and compare our results for the case where $\beta = 1.8$ (see Sec. \ref{sec:rda}), the value for dust in the galactic disk found by the \citet{planck2011} and consistent with other studies of nearby galaxies (\citealp{boselli2012}; \citealp{galametz2012}; Cortese et al. submitted). The fits were done by performing a $\chi^{2}$ minimisation using a simple gradient search method.

We estimate errors on the best-fit model parameters via a bootstrap technique. After the best-fit parameters are determined, 200 new sets of data points are created by selecting random flux values from within the errorbars of the observed fluxes. Each new dataset is then fitted again to find the alternative best fitting parameters. We then calculate the 68\% interval in both the upper and lower parameter distributions, setting these intervals as the new upper and lower limits. The differences in the parameter values of the original best-fit solution and the extreme values from the bootstrap technique are then taken as our uncertainties in the best-fit model parameters.

\subsection{Integrated dust properties}\label{sec:intphot}

Before proceeding with the resolved dust analysis, we first tested our methodology using integrated quantities. We determined the total flux densities in each FIR (i.e. 70 to 850~$\mu$m) waveband via aperture photometry utilizing \textsc{funcnts}, part of the \textsc{funtools} software for DS9. An elliptical source aperture centered on the galaxy was defined based on the extent of the 500~$\mu$m emission (shown in Fig.~\ref{fig:maps}). We use identical source apertures for all images except for the SCUBA 850~$\mu$m image, which covers a much smaller area than the other observations. To avoid the residual noise seen at the SCUBA field edges, we reduced the size of the source aperture used to measure the total galaxy emission in this band. 

Our PACS flux densities are 108 $\pm$ 8 Jy, 259 $\pm$ 13 Jy and 352 $\pm$ 28 Jy at 70, 100 and 160~$\mu$m respectively. These fluxes are consistent within the errors with the MIPS photometry presented by \citet{bendo2012b}, reported as 97 $\pm$ 10 Jy and 287 $\pm$ 35 Jy at 70 and 160~$\mu$m. Some of the variation between the MIPS and PACS fluxes may be the result of differences in the filter profiles. We successfully reproduce the SPIRE flux densities found by \cite*{bianchi2011}: $F_{250} = $ 155 $\pm$ 11 Jy, $F_{350} = $ 66 $\pm$ 6 Jy and  $F_{500} = $  23  $\pm$ 3 Jy. The discrepancies likely arise from a combination of differences in the choice of apertures in our photometry, updates to the flux calibrations and beam sizes, and sky subtraction. The SCUBA 850~$\mu$m image has a total flux density of 5 $\pm$ 1 Jy, consistent with values reported by \cite[][$F_{850} = $ 4.6$\pm$0.6 Jy]{alton1998} and \cite[][$F_{850} = $ 4.8$\pm$0.6 Jy]{israel1999}, who use an older calibration \citep[see also][]{haas2002}. Thus, our flux densities are consistent with those previously reported in the literature. These results are summarised in Table \ref{tab:fluxes}.
  
\begin{table}
 \centering
 \begin{minipage}{\columnwidth}
  \caption{Total flux densities and errors measured via aperture photometry for NGC~891.}
  \label{tab:fluxes}
  \begin{center}
  \begin{tabular}{c c}
\hline
\hline
$\lambda$  & Flux $\pm$ error  \\
 ($\mu$m) & (Jy) \\
\hline
70   &  108 $\pm$ 8\\
100  &  259 $\pm$ 13  \\
160  &  351 $\pm$ 28 \\
250  &  155 $\pm$ 11 \\
350  &  66 $\pm$ 6  \\
500  &  23 $\pm$ 3  \\
850  &   5  $\pm$ 1 \\
\hline
\end{tabular}
\end{center}
\end{minipage}
\end{table}

\begin{figure*}
\begin{center}
\includegraphics[width=1.0\textwidth]{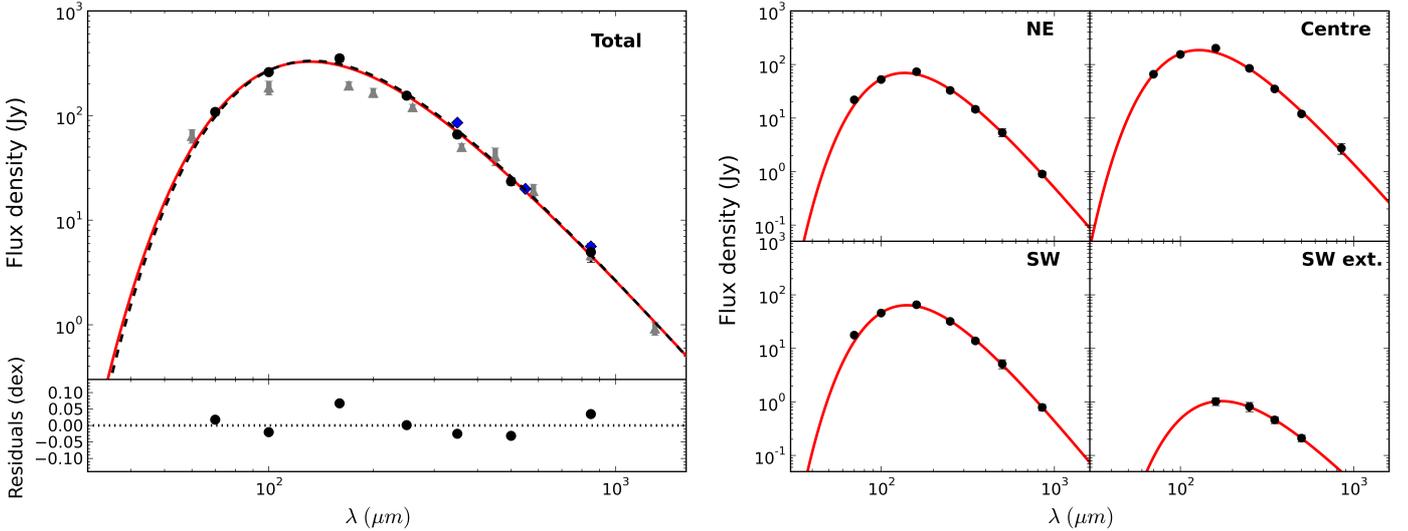}
\end{center}
\vspace{-0.25cm}
\caption[Integrated SEDs]{\textit{Left}: The total FIR/submm SED of NGC~891 obtained from the integrated PACS, SPIRE and SCUBA fluxes (black circles). The red line shows the best-fit one component model, where the 70~$\mu$m point is used as an upper limit. The black dashed line represents the best-fit one component model found when including the FIR/submm GAUFLUX fluxes reported in the \textit{Planck} HIFI Catalogue of Compact Sources (blue diamonds; \citealp{planck282013}). For comparison with previous FIR/submm experiments, the flux measurements from IRAS, ISO and IRAM presented in \citet[][see their Table 2]{popescu2011} are shown (grey triangles). \textit{Right}: SEDs measured for different locations in the galaxy, following \citet{dupac2003a} and \citet*{bianchi2011}. }\label{fig:intSED}
\end{figure*}

From these results, we construct the integrated SED of NGC~891 and fit the data with our one component modified black-body model, shown in Fig. \ref{fig:intSED}. For brevity, we focus our discussion on the SED fitting where we allow $\beta$ to vary as a free parameter, whilst also presenting the results derived from fixing $\beta$ = 1.8 in Table~\ref{tab:regions}. We find a dust mass and temperature of $\log M_{\mathrm{dust}} =$ 7.93 $\pm$ 0.05 M$_{\odot}$ and $T_{\mathrm{dust}} = 23.1 $ $\pm$ 1.2 K, with a $\beta = $ 1.77  $\pm$ 0.17. The total dust mass we uncover (8.5 $\times$ 10$^{7}$ M$_{\odot}$) is higher than those reported in most previous studies; past results have ranged from $M_{\mathrm{dust}} \sim$ 1.9 to 7 $\times$ 10$^{7}$ M$_{\odot}$ (\citealp{alton2000}; \citealp{popescu2004}; \citealp{galliano2008}). Our higher dust masses are likely due to some combination of the higher FIR fluxes measured by \textit{Herschel} compared to the observations from previous FIR experiments and the different assumptions for parameters used in the dust models of these studies.

We briefly note that NGC~891 appears in the \textit{Planck} HIFI Catalogue of Compact Sources \citep{planck282013}. The reported total flux densities measured at FIR/submm wavelengths, i.e. 350, 550 and 850~$\mu$m, using the photometry derived from Gaussian fitting, GAUFLUX, are $F_{350} =$ 85.5 $\pm$ 0.7 Jy, $F_{550} =$ 19.9 $\pm$ 0.3 Jy and $F_{850} =$ 5.6 $\pm$ 0.1 Jy. We find these flux values are consistent with those at the respective PACS/SPIRE wavebands. If we include these three data points in our SED fitting, we derive nearly equivalent dust properties as those stated earlier: $\log M_{\mathrm{dust}} =$ 7.96 $\pm$ 0.07 M$_{\odot}$, $T_{\mathrm{dust}} = 22.6 $ $\pm$ 1.4 K, with a $\beta = $ 1.80  $\pm$ 0.21 (see Fig~\ref{fig:intSED}). Thus, the difference in the best-fit models is negligible.

As a further check for consistency with the literature, we attempt to reproduce results reported in \cite{dupac2003a}. They traced the FIR SED using observations from ISO, IRAS/HiRes and PRONAOS. The SEDs were constructed from the flux density in three circular apertures with 1.5 arcmin radii (see Fig.~\ref{fig:maps}), where one is centred on the galaxy and other two placed at $\pm$ 3 arcmin along the galaxy major axis i.e. north-east (NE) and south-west (SW). Their one component modified blackbody fit to the SEDs yield dust temperatures of 23.5 K, 19.6 K and 18.1 K for the centre, NE and SW regions, respectively, with a range in $\beta \approx$ 1.41-1.96. We determine a similar dust temperature for the center region, but the NE and SW regions are warmer by approximately 1.5 to 2 K (see Fig.~\ref{fig:intSED} and Table~\ref{tab:regions}). Although our dust emissivities are slightly higher than \cite{dupac2003a}, ranging from  $\beta \approx$ 1.76-2.10, they are \textit{just} within the errors (less than $\pm$ 0.30). Such agreement in the results is encouraging when one considers that the two datasets came from different instrumentation. 

On the SW end of the disk, an asymmetric extension of the disk has been previously detected in ISOPHOT 170 and 200~$\mu$m images and found to correspond with the \hi~disk \citep{popescu2003}. The feature was also observed in the SPIRE 250, 350 and 500~$\mu$m maps by \cite*{bianchi2011}. They calculated the SW extension contained $M_{\mathrm{dust}} \sim$ 6 $\times$ 10$^{5}$ M$_{\odot}$, albeit with a factor of two error due to the difficulty in determining the dust temperature using only SPIRE flux densities. By adopting a \hi~mass of 2.5 $\times$ 10$^{8}$ M$_{\odot}$ in the SW extension and a dust-to-gas mass ratio of 0.006 (similar to the Milky Way and other galaxies, \citealp[e.g.,][]{draine2007}), \citet*{bianchi2011} further estimate a total dust mass of 1.5 $\times$ 10$^{6}$ M$_{\odot}$ for the extension. Combining the new PACS 70, 100 and 160~$\mu$m maps with the existing SPIRE observations, we can more accurately trace the peak of the integrated SED of the SW extension. We defined an aperture centred on $\alpha = 2^\mathrm{h}\ 22^\mathrm{m}\ 20\fs66$, $\delta = 42\degr\ 14\arcmin\ 56\farcs43$ (J2000.0) that included all the pixels with detections greater than 3-$\sigma$ in the 500~$\mu$m map (see Fig. 1 in \citealp*{bianchi2011}). We detect the SW extension in the PACS  160~$\mu$m map, however there is no detection of the feature in the 70 and 100~$\mu$m maps. Thus, we only include the PACS/SPIRE 160 to 500~$\mu$m data in the SED for this feature. The integrated SED and best-fit one component model of the SW extension is shown in Fig. \ref{fig:intSED}. We obtain a dust mass of 8.5 $\times$ 10$^{5}$ M$_{\odot}$, i.e. 1\% of our total dust mass estimate, lying between the two estimates of \citeauthor*{bianchi2011} and with an improved error of 15\%. However, the dust temperature and emissivity index have larger errors than typically produced from our method, most likely due to the existing difficulty in estimating the dust temperature without the constraint afforded by the 70 and 100~$\mu$m data. Yet, our results demonstrate consistency with the previous study.

Having tested our methodology via the reproduction of results from previous studies, we now focus on the central goal of this project: mapping the dust mass and temperature.

\begin{table}
 \centering
 \begin{minipage}{\columnwidth}
  \caption{The best fit parameters determined from fitting a model to the total SED, and to the SED observed in different regions.}
  \label{tab:regions}
  \begin{center}
  \begin{tabular}{l c c c c }
\hline
\hline
Aperture & $\log M_{\mathrm{dust}}$  & $T_{\mathrm{dust}}$  & $\beta$ & $\chi^{2}$ \\
  & (M$_{\odot}$)  & (K)  &   &   \\
\hline
$\beta$ free &   &    &  &  \\
\hline
 &   &    &  &  \\[0.025ex]
 Total & 7.93$^{+0.07}_{-0.04}$ &   23.1 $\pm$ 1.2  & 1.77 $\pm$ 0.17 & 0.94 \\[1.1ex]
Centre & $7.62^{+0.07}_{-0.03}$  &  23.8 $\pm$ 1.2  & 1.76 $\pm$ 0.17 & 1.50 \\[1.1ex]
NE  &  7.34$^{+0.07}_{-0.06}$  &  21.3 $\pm$ 1.0  & 2.10 $\pm$ 0.26 & 0.39 \\[1.1ex]
SW   & 7.37$^{+0.07}_{-0.07}$  &  20.2 $\pm$ 1.0  & 2.09 $\pm$ 0.16 & 0.21 \\[1.1ex]
SW ext.    & 5.93$^{+0.08}_{-0.08}$  &  20.2 $\pm$ 4.5  & 1.40 $\pm$ 0.80 &  0.48 \\[1.1ex]
\hline
$\beta =$ 1.8 &   &    &  &  \\
\hline
 &   &    &  &  \\[0.025ex]
 Total & 7.87$^{+0.03}_{-0.04}$ &   23.1 $\pm$ 0.3  & - & 0.97 \\[1.1ex]
Centre & 7.58$^{+0.02}_{-0.03}$  &  23.6 $\pm$ 0.3  & - & 1.51 \\[1.1ex]
NE  &  7.20$^{+0.02}_{-0.04}$  &  22.9 $\pm$ 0.3  & - & 0.72 \\[1.1ex]
SW   & 7.18$^{+0.02}_{-0.05}$  &  22.5 $\pm$ 0.3  & - & 1.08 \\[1.1ex]
SW ext.    & 5.66$^{+0.10}_{-0.07}$  &  21.6 $\pm$ 0.8  & - &  2.25 \\[1.1ex]
\hline
\end{tabular}
\end{center}
\end{minipage}
\end{table}

\subsection{Resolved dust analysis}\label{sec:rda}

\begin{figure}
\begin{center}
\includegraphics[width=0.85\columnwidth]{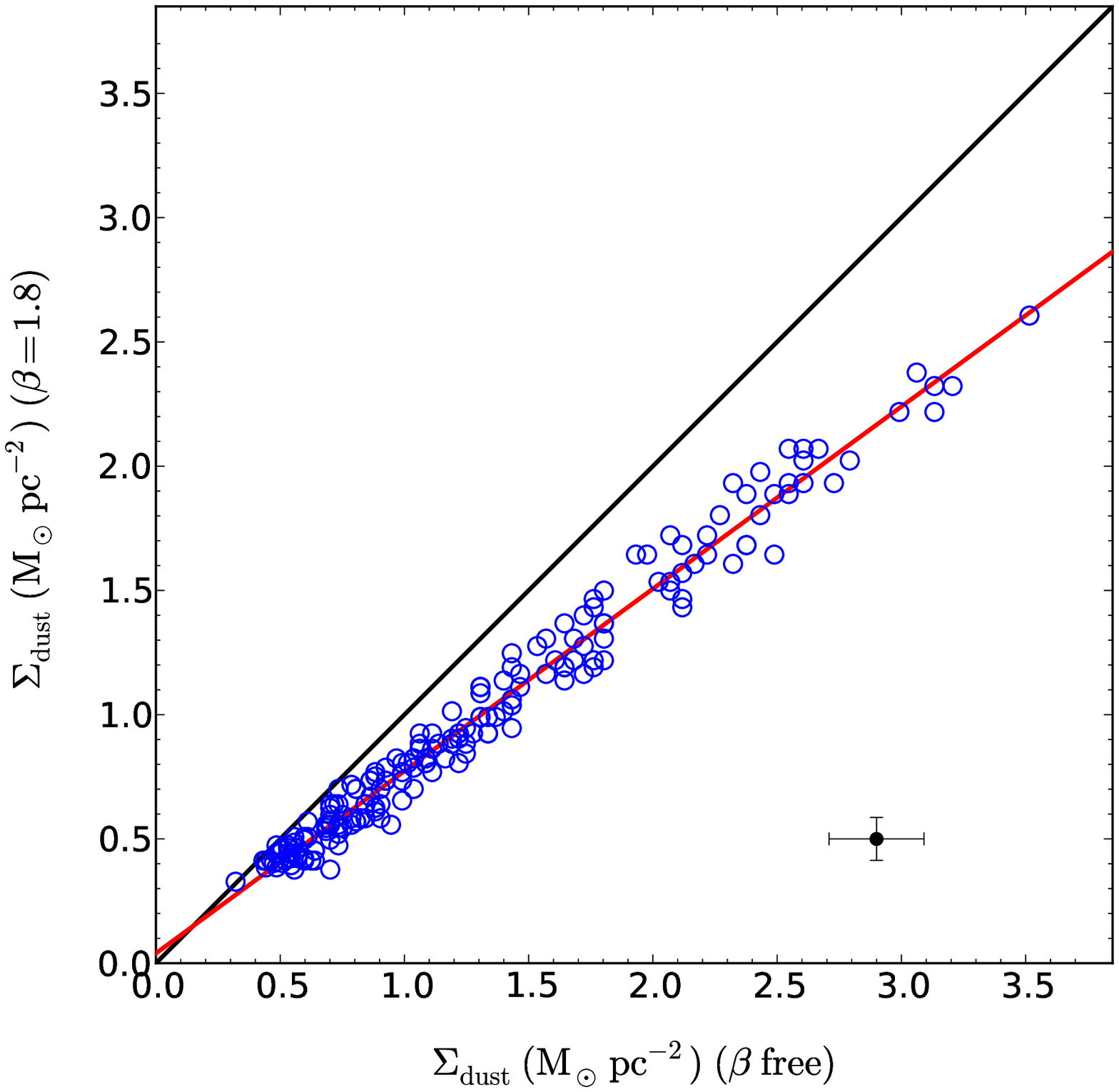}
\vspace{-0.5cm}
\includegraphics[width=0.85\columnwidth]{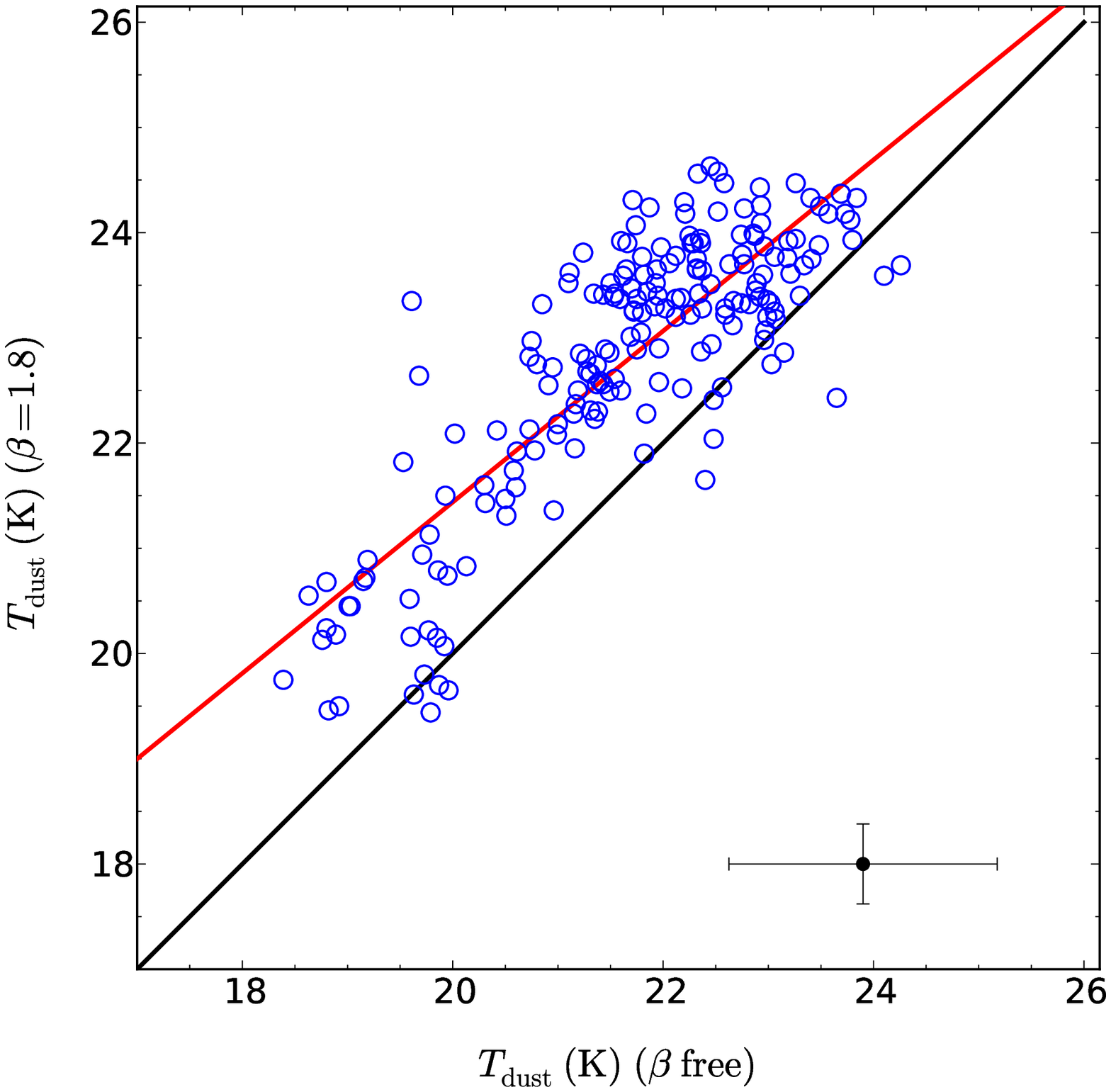}
\end{center}
\caption[Comparison of varying versus fixed dust emissivity]{A comparison of the dust masses (\textit{upper panel}) and  dust temperatures (\textit{lower panel}) obtained from allowing $\beta$ to vary versus fixing the $\beta$ value. The 1:1 relationship and our linear best fit relationship are indicated by the black and red lines, respectively.}\label{fig:betatemp}
\end{figure}

Using the PACS, SPIRE and SCUBA images, we fit the SEDs of each pixel in the image cube and construct maps of the dust properties. We only consider pixels with a detection greater than 5-$\sigma$ in all maps used for the SED fitting, yielding dust surface density and temperature estimates for 192 pixels. In Fig. \ref{fig:betatemp}, we compare the dust surface densities and temperatures derived from fitting the pixel SEDs using a modified blackbody with $\beta$ varying as a free parameter, against the results we obtain from fixing $\beta = 1.8$. For the derived dust masses, allowing $\beta$ to vary as a free parameter typically yields masses around 15 to 25\% higher than if we keep $\beta$ fixed; the median dust mass per pixel is 3.4~$\times$~10$^{5}$~M$_{\odot}$ for a varying $\beta$ and 2.5~$\times$~10$^{5}$~M$_{\odot}$ for $\beta = 1.8$ (corresponding to \sdust \ $=$ 1.1 and 0.8~M$_{\odot}$~pc$^{-2}$, respectively). The dust mass surface density shows good agreement between the two fitting methods, being offset from the 1:1 relationship with a mean difference of 0.32 M$_{\odot}$ pc$^{-2}$ and a small scatter of $\sigma$ = 0.20~M$_{\odot}$~pc$^{-2}$. The corresponding dust temperatures show similar agreement. Allowing $\beta$ to vary as a free parameter yields temperatures around 10\% lower than if we keep $\beta$ fixed. The median dust temperature is 21.7 K  for a varying $\beta$ and 23.2 K for $\beta = 1.8$. The scatter in the temperatures is greater than for dust masses, with a mean difference of 1.1 K and a $\sigma$ = 0.8 K. We examined the effect of varying the fixed value of the spectral index from $\beta = 1.8$ to $\beta = 2.0$, finding that the best agreement between the fixed and free $\beta$ values occurred at $\beta = 1.9$. In fact, this result merely reflects that $\beta = 1.9$ is the average value obtained when $\beta$ is allowed to vary as a free parameter (see also Fig. \ref{fig:mapsdustprops}). Thus, we find consistent results whether or not $\beta$ is a free or fixed value. 

\subsubsection{The $T_{\mathrm{dust}}$-$\beta$ relation}

The fitted dust emissivity index is a parameter that describes the slope and width of the blackbody curve and might, therefore, be related to the physical properties of the dust grains, including composition, grain size and equilibrium temperature. By fitting the observed FIR/submm SEDs of various sources with modified blackbody spectra, \citeauthor{dupac2001} (\citeyear{dupac2001}, \citeyear{dupac2002}, \citeyear{dupac2003b}) suggested that $\beta$  decreases with increasing dust temperatures. They found $\beta \sim 2$ in cooler regions ($T_{\mathrm{dust}} \approx 20$ K) and $\beta \sim 0.8-1.6$ for warmer environments ($T_{\mathrm{dust}} \approx 25-80$ K). This anticorrelation has since been observed in numerous experiments probing FIR/submm SEDs. ARCHEOPS data showed a stronger inverse relationship with $\beta$ ranging from 4 to 1 between $T_{\mathrm{dust}} \approx$ 7 and 27 K \citep{desert2008}, and a similar relation between $T_{\mathrm{dust}}$ and $\beta$ was found in BOOMERanG observations of eight clouds at high Galactic latitude \citep{veneziani2010}. Recent studies using \textit{Herschel} observations of the inner regions of the Galactic plane \citep{paradis2010} and Galactic cirrus emission \citep{bracco2011} also find an inverse relationship. 

There is some debate about the nature of such an inverse $T_{\mathrm{dust}}$-$\beta$ relationship, which may not have a physical origin, but rather arises from the data and/or methods used to derive the quantities (see e.g., \citealp{shetty2009a}; \citealp{foyle2012}; \citealp{juvela2013}). An inverse correlation between $T_{\mathrm{dust}}$ and $\beta$ has been shown to arise from least-squares fits due to the uncertainties in data \citep{shetty2009a} and line-of-sight temperature variations \citep{shetty2009b}, since a one-component modified blackbody may not be appropriate for modelling the emission from dust that has a range of temperatures. Such line-of-sight temperature variations may lead to an underestimation of $\beta$ (e.g., \citealp{malinen2011}; \citealp{juvela2012}). Furthermore, spurious temperature variations may be introduced, such as peaks in the temperature distribution where there are no identifiable heating sources \citep{galametz2012}.  

Yet in a pixel-by-pixel analysis of dust and gas in the Andromeda galaxy, \cite{smith2012} observed a radial dependency on the $T_{\mathrm{dust}}$-$\beta$ relationship; pixels at radii $R > 3.1$ kpc produce a steeper relation compared to pixels within the radius $R < 3.1$ kpc. They argue that neither artefacts in the fitting due to noise nor the need to account for line-of-sight temperature variations can explain this clear separation between the relations from the inner and outer regions of Andromeda. Furthermore, the relation from \citeauthor{smith2012} for the outer region ($R > 3.1$ kpc) is in good agreement with the $T_{\mathrm{dust}}$-$\beta$ relationship observed in the global dust properties derived for galaxies in the \textit{Herschel} Virgo Cluster Survey \citep{davies2013}. These results suggest a possible physical origin of the relationship.

\begin{figure}
\begin{center}
\includegraphics[width=0.99\columnwidth]{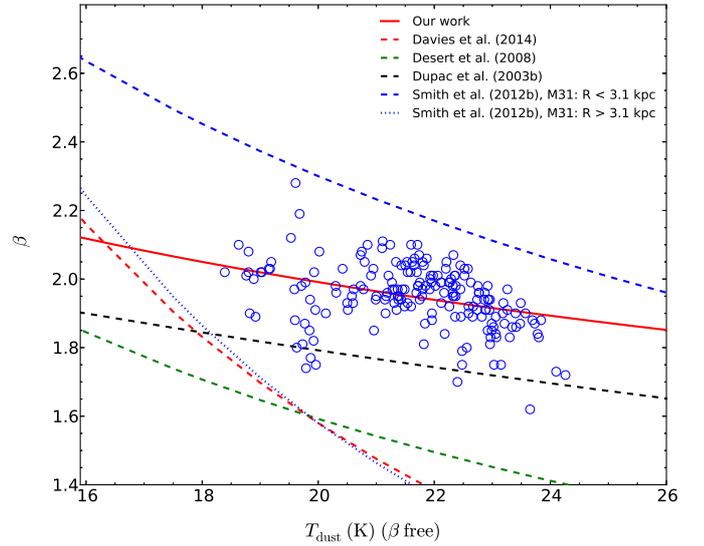}
\end{center}
\vspace{-0.4cm}
\caption[The relationship between the dust spectral index and temperature]{The anticorrelation we observe between $\beta$ and dust temperature from fitting a one component modified blackbody to the pixel-by-pixel SEDs constructed from our FIR/submm images.}\label{fig:betatemprel}
\end{figure}

Plotting our results (see Fig. \ref{fig:betatemprel}), we observe a similar relationship between dust emissivity index and temperature. A weak negative correlation exists between the two quantities with a Spearman coefficient of rank correlation of $\rho$ = -0.52, corresponding to a probability P($\rho$) $>$ 99.9\% that the two variables are correlated. Adopting the same parametrisation for the relation as in previous works (e.g., \citealp{smith2012}), we find a best fit relationship of
\begin{eqnarray}
\beta = 1.99\ \left(\frac{T_{\mathrm{dust}}}{20\ \mathrm{K}}\right)^{-0.29}
\end{eqnarray}
The distribution of the points are fairly consistent with previous studies, lying offset from the \cite{dupac2003b} relation and between the two radial relations of \cite{smith2012}, although their best-fit relation is steeper at lower temperatures compared to our best fit. 

As NGC~891 is a nearly fully edge-on spiral ($i >$ 89$^{\circ}$, e.g., \citealp*{kregel2005}), we cannot investigate the existence in NGC~891 of a radial break in the $T_{\mathrm{dust}}$-$\beta$ relation as observed in Andromeda. We attempt to exploit the inclination of NGC~891 by examining any dependence of the $T_{\mathrm{dust}}$-$\beta$ relation on the vertical height from the mid-plane. We roughly divide the pixels covering the galaxy into two bins: one containing all pixels with a perpendicular distance from the mid-plane greater than $\pm$ 1 kpc, and one containing all pixels located within $\pm$ 1 kpc from the mid-plane. This distance from the disk was chosen based on the extent of the molecular gas in the disk, which has been shown to have a $\pm$ 1 kpc vertical extension. We remind the reader that at a distance of 9.6 Mpc each 12$\arcsec$ pixel corresponds to approximately 558 pc and, due to the PSF of the SPIRE 500~$\mu$m image of 36\arcsec, are not independent. We do not find a significant offset in the mean $\beta$ of these two regions. The $T_{\mathrm{dust}}$-$\beta$ relationships also do not show significant variance between the two bins. This is evident when one inspects the dust temperature and spectral index maps presented in Fig. \ref{fig:mapsdustprops}. The temperature distribution is clumpy and asymmetric about the mid-plane, whereas $\beta$ has a much smoother distribution. 

Thus, we do not find a statistical variation of the $T_{\mathrm{dust}}$-$\beta$ relation on the vertical height from the disk, suggesting that any change in the $T_{\mathrm{dust}}$-$\beta$ relation is predominantly linked to radial variations from the interior to the exterior of a galaxy \citep{smith2012}. However, since we are fitting a one-component modified blackbody to dust emission integrated along the line-of-sight through the galaxy disk, covering a range of environments and temperatures, we stress that such a physical interpretation for the variation of the relationship should be regarded with caution.

\subsubsection{Spatial distributions}

Given the plethora of possible problems and caveats that must be considered when attempting to interpret the fitted dust emissivity index, many of which we have discussed above, and, since we find consistent dust properties whether or not $\beta$ is a free variable or a fixed value, we decide to adopt the dust properties derived from fixing $\beta =$ 1.8 for the remainder of our analysis. 

\begin{figure*}
\begin{center}
\includegraphics[width=0.99\textwidth]{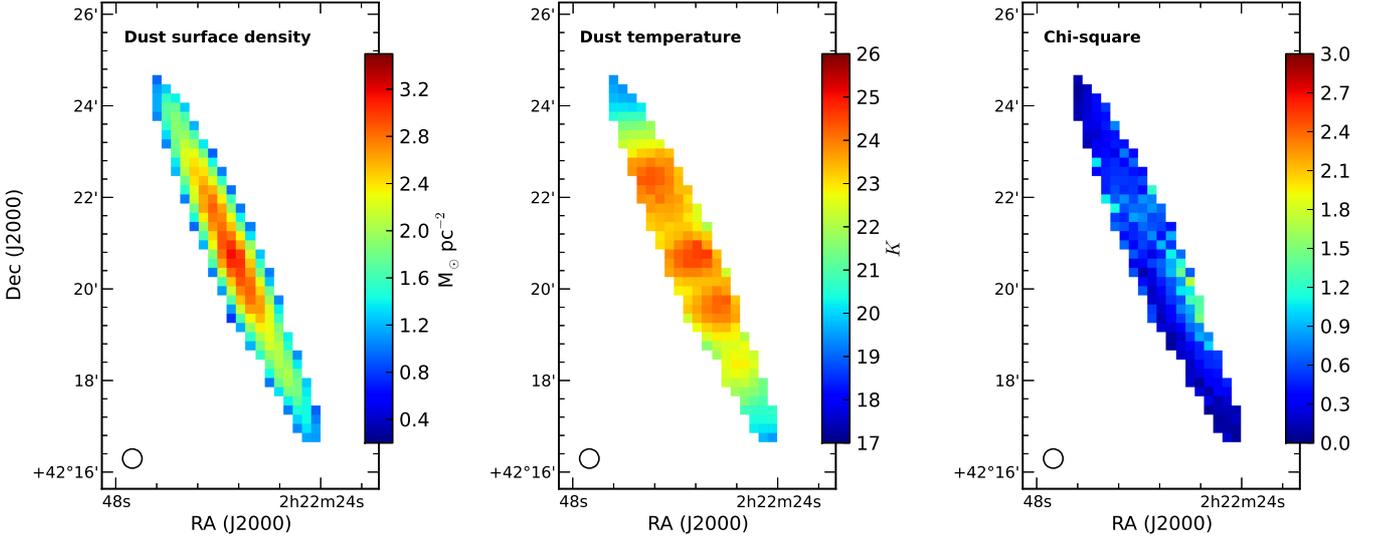}
\end{center}
\vspace{-0.5cm}
\caption[Dust property maps]{Maps of the dust mass surface density (\textit{left}), dust temperature (\textit{middle}) and $\chi^{2}$ value (\textit{right}), obtained from fitting a one component, modified blackbody model with $\beta =$ 1.8 to the SED of each pixel. The maps are centred on $\alpha = 2^\mathrm{h}\ 22^\mathrm{m}\ 33\fs0$, $\delta = 42\degr\ 20\arcmin\ 57\farcs2$ (J2000.0), with the beam size indicated by the black circle. The pixel size is 12$\arcsec$ pix$^{-1}$ corresponding to 0.56 kpc at a distance of 9.6 Mpc. }\label{fig:mapsdustprops}
\end{figure*}

The resulting maps are presented in Fig.~\ref{fig:mapsdustprops}. The mean dust mass surface density is 1.0 M$_{\odot}$ pc$^{-2}$, which peaks at 2.6 M$_{\odot}$ pc$^{-2}$ in the centre. Either side of the centre and along the semimajor axis are two smaller local maxima that coincide with the surface brightness peaks seen in the WISE, PACS, SPIRE and SCUBA images (see Fig.~\ref{fig:maps}). The distribution shows a fairly smooth decrease away from the galaxy mid-plane. However, the dust temperature has a much more uneven distribution. Ranging from $\approx$ 17 to 24 K with a median of 22.9 K, the temperature peaks in the centre and again in a region on the NE end of the disk. The dust in this region appears warmer than the average dust temperature on the opposite end of the disk i.e., towards the SW, an observation which is most prominent in the averaged radial profile of the dust temperature map (see Fig.~\ref{fig:radialprofs}). In fact, the dust temperature profile reflects the same features and asymmetry as seen in the radial profile of the 24~$\mu$m~emission. This asymmetry in the disk has been previously noted in the literature (e.g., \citealp{kamphuis2007}), and we shall discuss this behaviour in detail in Section \ref{sec:diskasym}. We briefly note that the temperature profile exhibits the same break as the FIR/submm profiles, indicating that the break in the radial profiles is most likely due to a change in the ISM properties at radii $\pm$ 12 kpc. Returning to Fig.~\ref{fig:mapsdustprops}, we also show the $\chi^{2}$ value used to fit the SED of each pixel. The majority of the pixel SEDs have $\chi^{2}$ values $<$ 1.5. Those pixels with $\chi^{2} >$ 1.5 are typically at the outskirts of the disk, where the flux density from the SCUBA 850~$\mu$m are less certain with higher errors, as confirmed from a visual inspection of the individual SEDs for these pixels.

Furthermore, we investigated whether the low resolution of the SPIRE 500~$\mu$m data (36$\arcsec$) affected our results by omitting this data and performing the same pixel-by-pixel SED fitting at the resolution of the SPIRE 350~$\mu$m data (24$\arcsec$). We find that increasing the resolution yields the same distribution of dust properties, albeit with slightly higher mass surface densities and lower temperatures ($\sim$ 6 \%, i.e. much smaller than the differences seen for treating $\beta$ as a fixed or variable quantity). However, the SEDs of some individual pixels are less constrained on the Rayleigh-Jeans side, mainly due to the sensitivity of the fitting to the 850~$\mu$m flux with respect to the overall FIR SED (see Fig. \ref{fig:intSED}). Since the pixel-by-pixel fits to the SED are better constrained when including rather than excluding the 500~$\mu$m data and the enhanced resolution neither improves nor affects our results, we continue the analysis including the SPIRE 500~$\mu$m data at the lower resolution.

\section{Gas and dust}\label{sec:dustgas}

Having determined the dust properties of NGC~891 in the previous section, we now combine this new information with measurements of the gas content and examine the relationships between gas and dust in the disk.

\subsection{Estimating gas content}

We first calculate the masses and corresponding surface densities of the various gas components. We base our analysis primarily on the pixels covered by our CO(3-2) map, since it covers a much larger fraction of the galaxy compared to the available \citet{scoville1993} CO(1-0) emission map (see Fig. \ref{fig:maps}). We include the latter CO(1-0) observations only as an additional test of the results of our analysis.

\subsubsection{CO(3-2) observations}

We derive a molecular gas mass estimate for NGC 891 from the new map of $^{12}$CO(J=3-2) emission. For all detected pixels, the integrated CO(3-2) line intensity is converted to a H$_{2}$ mass, $M_{\mathrm{H}_{2}}$, according to the equation
\begin{eqnarray}\label{eqn:coconv}
M_{\mathrm{H}_{2}} = A\,m_{\mathrm{H}_{2}}\,\left(\frac{X_{CO}}{1\times 10^{20} \frac{\mathrm{cm}^{-2}}{\mathrm{K\ km s}^{-1}} }  \right)  \frac{I_{\mathrm{CO(3-2)}}}{\eta_{\mathrm{mb}}\left(\frac{I_{\mathrm{CO(3-2)}}}{I_{\mathrm{CO(1-0)}}}\right)} 
\end{eqnarray}
where $I_{\mathrm{CO(3-2)}}$ is the total integrated line intensity expressed in units of K~km~s$^{-1}$, $A$ represents the surface of the CO(3-2) emitting region and $m_{\mathrm{H}_{2}}$ is the mass of a hydrogen molecule. The scaling factor to convert an antenna temperature $T_{\mathrm{A}}$ into a main beam temperature $T_{\mathrm{mb}}$ at the JCMT is $\eta_{\mathrm{mb}} =$ 0.6. We assume a value of 0.3 for the CO(3-2)-to-CO(1-0) line intensity ratio corresponding to the typical ratios found in the diffuse ISM of other nearby galaxies \citep{wilson2009}. The CO-to-H$_{2}$ conversion factor, $X_{CO}$, has been difficult to constrain and varies significantly over a range of environments and metallicities (see \citealp{bolatto2013}, and references therein). Although we lack the observations to determine the metallicity gradient in NGC 891, we refrain from using an average metallicity gradient derived from observations of other nearby face-on spiral galaxies to estimate the radial variation in the $X_{CO}$ factor (similar to an approach used in \citealp{alton2000}) in order to avoid (i) the large uncertainties in applying such a gradient, and (ii) the uncertainties in the choice of radial relationship between $X_{CO}$ and metallicity (see e.g. Fig. 9 in \citealp{bolatto2013}). We therefore choose to adopt $X_{CO} = 2\,\times$ 10$^{20}$ cm$^{-2}$ [K km s$^{-1}$]$^{-1}$, based on the recommendation of \citet{bolatto2013} as a conservative choice for the disks of normal, solar metallicity galaxies and the Milky Way (see also \citealp{sandstrom2013}). Our adopted values result in a total H$_{2}$ mass of 2.9 $\times 10^{9}$ M$_{\odot}$ contained in the observed area. Lower estimates were obtained by \citet[][1.5 $\times 10^{9}$ M$_{\odot}$]{guelin1993} and \citet[][1.4 $\times 10^{9}$ M$_{\odot}$]{israel1999}. Our corresponding mean H$_{2}$ mass surface density is $\Sigma_{\mathrm{H}_{2}} =$ 67 M$_{\odot}$ pc$^{-2}$, where \shii \ ranges from 11 to 144 M$_{\odot}$ pc$^{-2}$. 

\subsubsection{\hi \ observations}

We estimate the \hi \ surface density using the total column density map from \citet{oosterloo2007}, finding a total integrated \hi \ mass of $M_{\mathrm{H}\tiny{\textsc{i}}} = 4.2 \times 10^{9}$ M$_{\odot}$ at our adopted distance of 9.6 Mpc. Previous estimates have ranged from 2.5 - 8 $\times 10^{9}$ M$_{\odot}$ (\citealp*{sancisi1979}; \citealp{rupen1991}; \citealp{guelin1993}). It has been shown that up to 30\% \ of the total mass is contained in a large gaseous halo extending $\sim$ 22 kpc from the disk (\citealp{swaters1997}; \citealp{oosterloo2007}). We calculate the disk contains $2.5 \times 10^{9}$ M$_{\odot}$. The median \hi \ mass surface density of the 12$\arcsec$ pixels is $\Sigma_{\mathrm{H}\tiny{\textsc{i}}} =$ 42 M$_{\odot}$ pc$^{-2}$ and ranges from 22 to 58 M$_{\odot}$ pc$^{-2}$ in the disk. The errors are estimated at 10\%. Furthermore, we note that we have not accounted for the effects of \hi \ self-shielding. Whilst we adopt the usual assumption of negligible self-opacity for the transition, \citet{braun2009} warn that this assumption may overlook the presence of hidden \hi \ features that are self-opaque in the 21 cm transition and thus underestimate the global atomic gas mass by up to 34\%$\pm$5\% (see also \citealp{braun2012}). However, any attempt of applying local opacity corrections to our column density map is not straightforward because (i) line-of-sight confusion will tend to overlap discrete features in both position and velocity, and (ii) the physical scales probed in our analysis ($\sim$500 pc) are considerably larger than the physical scales of any hidden \hi \ features ($\sim$50-100 pc). We therefore take a conservative approach and refrain from applying this correction, yet caution that our \hi \ surface density estimates may be underestimated by $\sim$30\% and continue our analysis with this caveat in mind.

\subsubsection{Total gas content}

From these measurements of the atomic and molecular hydrogen masses, we estimate the total gas mass in each pixel as $M_{\mathrm{gas}} = M_{\mathrm{H}\tiny{\textsc{i}}} + M_{\mathrm{H}_{2}}$ and include the contribution of helium and heavy elements (i.e., a factor 1.36). Thus, our integrated gas mass for NGC~891 is $M_{\mathrm{gas}} =$ 1.1 $\times 10^{10}$ M$_{\odot}$. For comparison, previous studies found total gas masses ranging from $M_{\mathrm{gas}} \approx$ 0.3-1.2 $\times$ 10$^{10}$ M$_{\odot}$ (see e.g., \citealp{guelin1993}; \citealp{israel1999}; \citealp{dupac2003a}). We briefly note that our integrated dust, H$_{2}$ and \hi \ masses place NGC~891 very close to the best-fit $M_{\mathrm{H}_{2}}$-$M_{\mathrm{dust}}$ and $M_{\mathrm{gas}}$-$M_{\mathrm{dust}}$ relations derived from late-type Virgo cluster galaxies \citep[see][]{corbelli2012}. The total gas content, $\Sigma_{\mathrm{gas}}$, has a mean surface density of 149 M$_{\odot}$ pc$^{-2}$ and ranges from 56 to 267 M$_{\odot}$ pc$^{-2}$.

\begin{figure*}
\begin{center}
\includegraphics[width=0.95\textwidth]{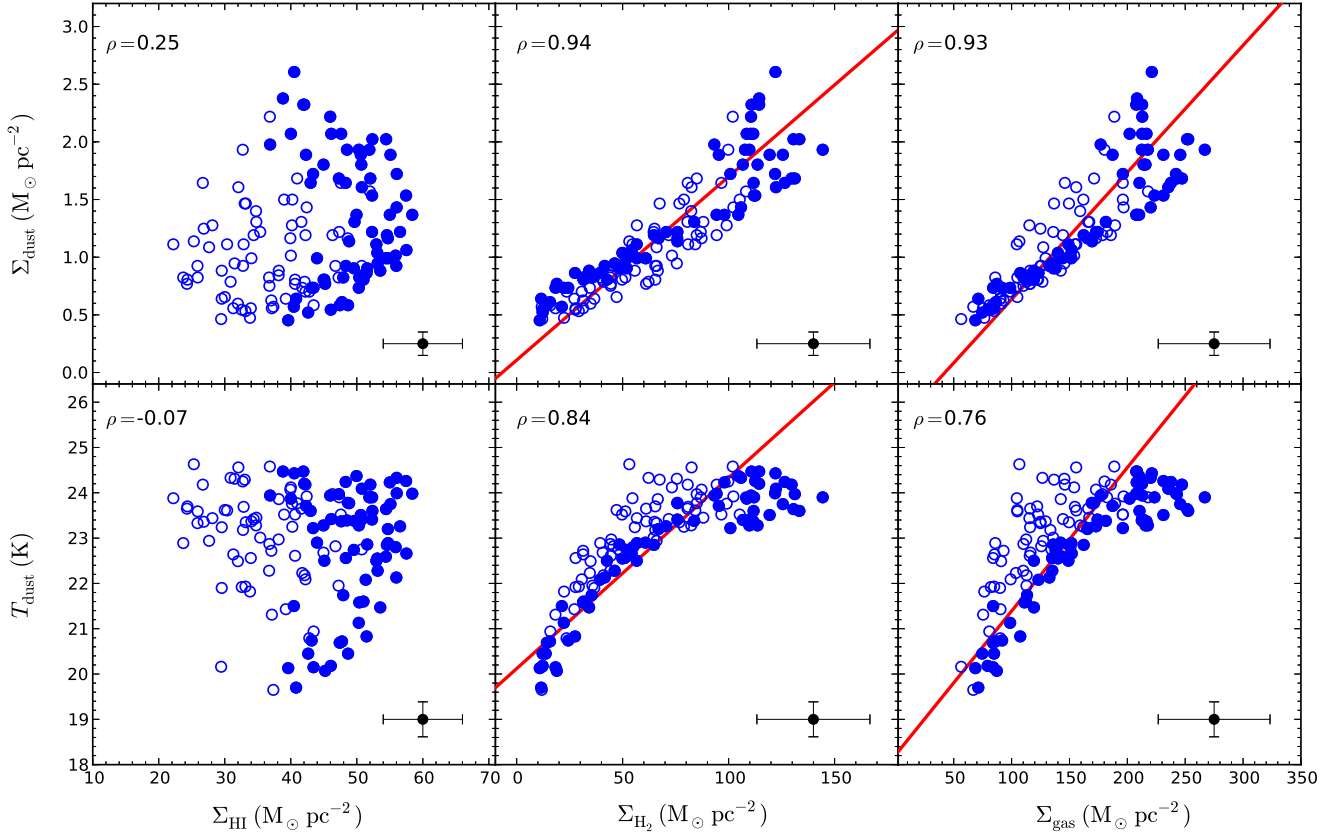}
\end{center}
\vspace{-0.3cm}
\caption[Relationships between the dust and gas]{The pixel-by-pixel relationships between the derived dust properties, \sdust \ (\textit{upper panels}) and $T_{\mathrm{dust}}$ (\textit{lower panels}), and the surface densities of the gaseous ISM components: \shi \ (\textit{left}), \shii \ (\textit{middle}) and \sgas \ (\textit{right}). We differentiate between on- and off-axis pixels using solid and open circles, respectively. The best linear fits are indicated by the red solid lines. Error bars are calculated from the standard deviations of the errors on individual pixels and the $\rho$ values are the corresponding Spearman correlation coefficients.}\label{fig:dustgasrelations}
\end{figure*}

\subsection{The dust-gas connection}

We now combine all our surface density estimates to study the relationships between dust and gas in the galaxy. We stress that all values related to the gas content reported henceforth in our analysis are derived from H$_{2}$ masses from the CO(3-2) map. 

We find several strong correlations between the pixel-to-pixel dust and gas surface densities. In Fig.~\ref{fig:dustgasrelations}, the derived dust surface density and temperature for each pixel are plotted against the corresponding \hi , H$_{2}$ and total gas surface densities. As expected, we find only a very weak correlation between \shi  \ and \sdust , as evident both in Fig. \ref{fig:dustgasrelations} and indicated by a low Spearman coefficient of rank correlation ($\rho$ = 0.25). We remind the reader that this is only true for the inner part of the galaxy. However, the clear correlation between \shii  \ and \sdust \ has a Spearman coefficient of $\rho$ = 0.94, corresponding to a probability P($\rho$) $>$ 99.9\% that the two variables are correlated. Interestingly, we find a similar correlation when combining the \hi \ and H$_{2}$ gas content; the correlation between \sgas \ and \sdust \ has a Spearman coefficient of $\rho$ = 0.93, with a corresponding P($\rho$) $>$ 99.9\%. Both the \shii -\sdust \ and \sgas -\sdust \ correlations are best fit with the linear relations
\begin{eqnarray}\label{eqn:h2dust}
\Sigma_{\mathrm{H}_{2}} = 62.8 \pm 0.6 \times \Sigma_{\mathrm{dust}} - 6.7 \pm 0.6\mathrm{, \ and}
\end{eqnarray}
\begin{eqnarray}\label{eqn:gasdust}
\Sigma_{\mathrm{gas}} = 90.8 \pm 0.7 \times \Sigma_{\mathrm{dust}} +42.4 \pm 0.9,
\end{eqnarray}
where all quantities are in units of M$_{\odot}$ pc$^{-2}$. Such correlations between \sdust \ and \shii \ are predicted from the physics of CO formation and dissociation and have already been observed in a number of environments. For example, \citet{romanduval2010} found linear correlations between \sgas \ and \sdust \ on even smaller spatial scales, via \textit{Herschel} observations of two molecular clouds in the LMC (see also \citealp{meixner2010}; \citealp{leroy2011}; \citealp{galliano2011}, see their Fig. 12). The \sgas -\sdust \ correlations have been reproduced in the simulated disk galaxies of \citet{bekki2013}, using star formation histories regulated by the time evolution of interstellar dust, the properties of which control H$_{2}$ formation rates. The dust surface density in the simulated disks correlated with \sgas \ than \shii , consistent with our observations. 

An immediate observation from these \shii - and \sgas -\sdust \ correlations is that they appear to become increasingly bifurcated with increasing surface density. Such bifurcation could indicate problems with offsets between images, misaligned position angles or a mismatching of the PSFs adopted for the image convolution. However, our investigations into the effects of misalignments between the images, by systematically introducing artificial offsets in the position angles and world coordinate systems of the images, failed to confirm that any of these issues were causing the bifurcation. In fact, these artificial misalignments tended to drastically increase the scatter of the observed correlations and thereby hide the bifurcation. The minimum scatter in the correlations actually occurs when adopting the position angle of 22.9$^{\circ}$, consistent with our conclusions when extracting the vertical and radial profiles (see Section~\ref{sec:submmmorph}). Taking the best fit lines to the \sgas -\sdust \ correlations (given by Equations~\ref{eqn:h2dust} and~\ref{eqn:gasdust}) as references, we note that the highest  \shii \ and \sgas \ pixels have lower \sdust \ values than these best fit lines would suggest, whereas the pixels with the highest \sdust \ tend to have lower \shii \ and \sgas \ values. In other words, there is a clear difference in the gas-to-dust ratios and these results imply that the gas-to-dust ratios in the galaxy center are much lower than the rest of the disk.  

\begin{figure*}
\begin{center}
\includegraphics[width=0.99\textwidth]{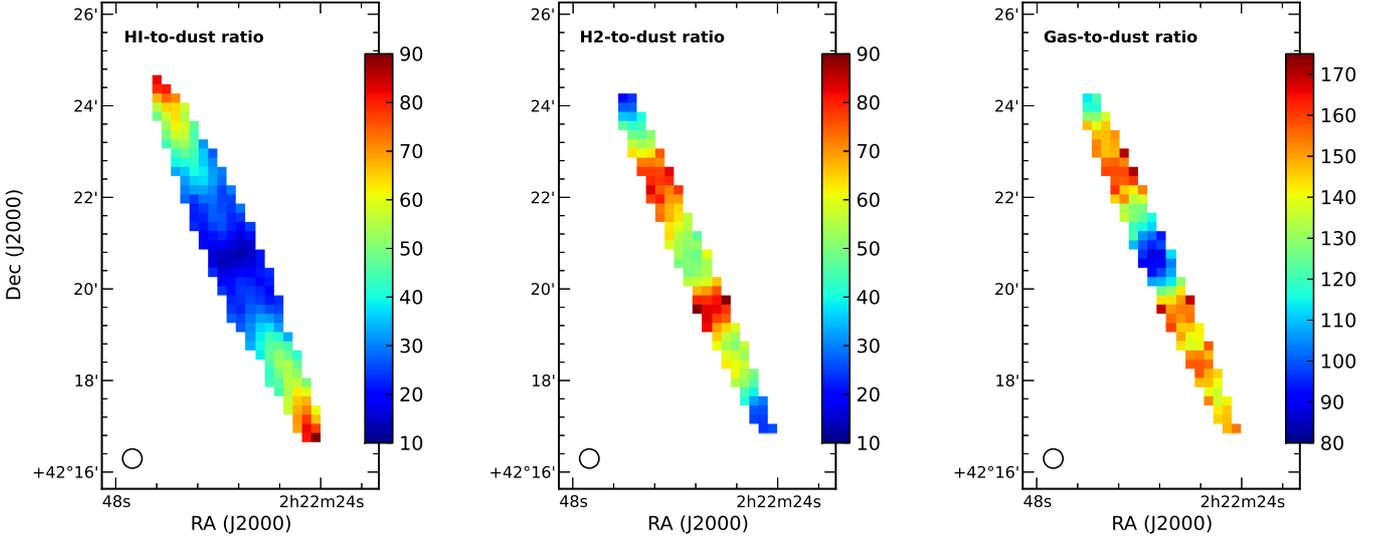}
\end{center}
\vspace{-0.5cm}
\caption[Dust property maps]{Maps of the \hi -to-dust ratio (\textit{left}), H$_{2}$-to-dust ratio (\textit{middle}) and the total gas-to-dust ratio (\textit{right}). The maps are centred on $\alpha = 2^\mathrm{h}\ 22^\mathrm{m}\ 33\fs0$, $\delta = 42\degr\ 20\arcmin\ 57\farcs2$ (J2000.0), with the beam size indicated by the black circle. The pixel size is 12$\arcsec$ pix$^{-1}$ corresponding to 0.56 kpc at a distance of 9.6 Mpc. }\label{fig:mapsgastodust}
\end{figure*}

To further explore the origin of this birfucation, Fig.~\ref{fig:mapsgastodust} presents the mapped ratios of the various gas surface densities to the dust surface density, i.e. the  \hi -to-dust ratio (\shi /\sdust), H$_{2}$-to-dust ratio (\shii /\sdust), and the total gas-to-dust ratio (\sgas / \sdust). The  \hi -to-dust ratio is lowest in the centre of the disk and increases with increasing radius, reflecting the observations that the dust surface density peaks in the centre whereas the \hi \ surface density dominates at larger radii. The lack of a clear radial trend lends further evidence to the lack of an overall relationship between \shi \ and  \sdust \ in the disk (Fig. \ref{fig:dustgasrelations}). In contrast, the H$_{2}$-to-dust ratio is fairly low (\shii /\sdust $\sim$ 50) at the disk centre, as predicted from the bifurcation of the \sgas -\sdust \ correlations, but peaks in two regions towards the NE and SW of the disk. Most intriguing is the spatial distribution of the total gas-to-dust ratio. The mean \sgas /\sdust \ is $\sim$ 140, slightly lower than the Galactic \sgas /\sdust \ value of 158 from \citet{zubko2004}. Most of the disk possesses a \sgas /\sdust \ ratio between 110 and 150, but there are two off-centre peaks of \sgas /\sdust \ $\approx$ 160 situated on each side of the disk at approximately $\pm$ 5 kpc from the center. These peaks also roughly coincides with the aforementioned peaks in the dust temperature distribution (see Fig. \ref{fig:mapsdustprops}). We briefly note a good agreement between the spatial variation of the gas-to-dust ratio observed in our maps and the variations inferred from the radial gas-to-dust profiles presented in \citet{alton2000}.

\begin{figure}
\begin{center}
\includegraphics[width=0.85\columnwidth]{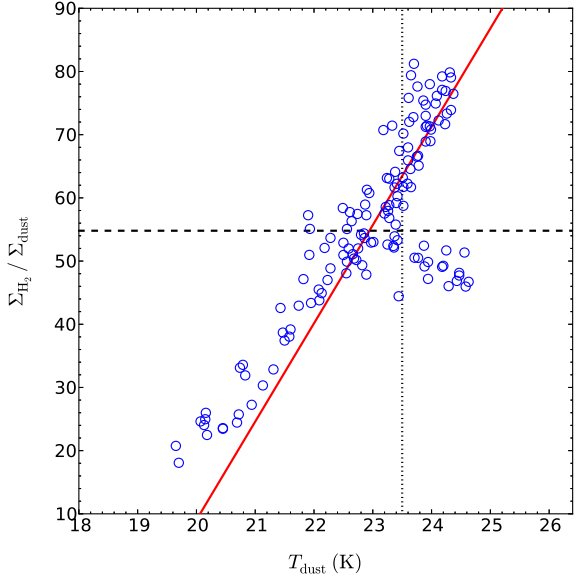}
\end{center}
\vspace{-0.3cm}
\caption[Dust temperature versus the H2-to-dust ratio]{The correlation between the H$_{2}$-to-dust ratio and the dust temperature. The best linear fits are indicated by the red solid lines, with the mean \shii / \sdust \ ratio represented by the dashed line. The cluster of points with \shii / \sdust \ ratio lower than the mean and possessing a higher dust temperature ($T_{\mathrm{dust}} > 23.5$ K, shown by the dotted line) are all central pixels.}\label{fig:gdrtemp}
\end{figure}

The dip in the gas-to-dust ratio in the galaxy center becomes particularly obvious when we consider the relationship between \sgas /\sdust \ and the dust temperature (see Fig.~\ref{fig:gdrtemp}). We find a tight relationship between these quantities for most of the disk, but the central pixels show a clear separation from this relation due to their higher dust temperatures but on average lower \sgas /\sdust \ values. The exact cause of this is uncertain, but may be indicating that we are underestimating the H$_{2}$ surface density. Whilst the assumption of a single value for the CO-to-H$_{2}$ conversion factor may be a conservative choice for the disks of normal, solar metallicity galaxies and the Milky Way, and so an appropriate choice for the majority of the disk of NGC 891, in this case it may not be appropriate for the nuclear regions. Though the $X_{CO}$ factor is observed to drop sharply in the central, bright regions of some galaxies, often coincident with bright CO emission and high stellar surface density (see e.g. \citealp{bolatto2013}; \citealp{sandstrom2013}), our choice of $X_{CO}$ factor may still be underestimating the \shii \ in the center. Another plausible scenario is that the CO(3-2) emission fails to trace the full molecular content in the densest central regions, particularly when integrating along the line-of-sight through the nucleus of an edge-on galaxy, as the \mbox{CO(3-2)} line could become optically thick in the centre of galaxies. The CO(3-2) line becomes optically thick close to a molecular hydrogen column density $N_{H_{2}} = 2 \times 10^{5}$ cm$^{-3}$, which corresponds to a cloud with $N_{CO} > 2 \times 10^{17}$ cm$^{-2}$ or $Tk\ >$ 550 K (see e.g. \citealp{imai2012}). Furthermore, the geometry of the molecular gas may play a role, as a clumpy distribution of molecular clouds will be locally optically thick and may therefore inefficiently contribute to the CO emission. Undetected CO emission in denser clumps may also explain why \citet{sakamoto1997} tend to find moderate H$_{2}$ densities from the observed CO emission.

Whilst the H$_{2}$-to-dust and gas-to-dust ratios show some spatial correlation with the dust temperature, evident from the coincidence of the peaks in their distributions (see also Fig.~\ref{fig:gdrtemp}), the gas surface densities typically show weaker correlations with dust temperature compared to the dust surface density (see Fig.~\ref{fig:dustgasrelations}). Weaker correlations are found between dust temperature and the H$_{2}$ and total gas surface densities than compared to the similar relations with dust surface density. The observed correlation between $T_{\mathrm{dust}}$ and \shii \ has a Spearman coefficient of rank correlation of $\rho$ = 0.84, corresponding to a probability of correlation of P($\rho$) $>$ 99.9\%. The $T_{\mathrm{dust}}$-\sgas \ relation has a Spearman coefficient of $\rho$ = 0.76, also with P($\rho$) $>$ 99.9\%. The best-fit linear solutions are given by
\begin{eqnarray}
\Sigma_{\mathrm{H}_{2}}  = 23.7 \pm 0.1 \times T_{\mathrm{dust}} - 477.7 \pm 2.3 \ \mathrm{, \ and}
\end{eqnarray}
\begin{eqnarray}
\Sigma_{\mathrm{gas}} = 31.5 \pm 0.2 \times T_{\mathrm{dust}} - 575.4 \pm 4.0,
\end{eqnarray}
where \shii \ and \sgas \ are in units of M$_{\odot}$ pc$^{-2}$ and $T_{\mathrm{dust}}$ is in K. Although we fit linear relations to the correlations, it is evident that such linear relationships are not really appropriate. For pixels containing $\Sigma_{\mathrm{H}_{2}} < $ 100 M$_{\odot}$ pc$^{-2}$, the plotted data appears to follow a linear correlation. This also seems to be reflected in the $T_{\mathrm{dust}}$-\sgas \ relation for pixels \sgas \ $< $ 200 M$_{\odot}$ pc$^{-2}$. However, it is not clear at present what causes the pixels at $\Sigma_{\mathrm{H}_{2}} > $ 100 M$_{\odot}$ pc$^{-2}$ (corresponding to \sgas \ $> $ 200 M$_{\odot}$ pc$^{-2}$) to flatten out towards a peak temperature, rather than continue to follow the linear trend found for the rest of the disk. One potential clue may be gained from the location of these pixels in the galaxy. 

Since we have shown that (1) the dust mass surface density typically decreases as a function of radius and $|z|$ from the galaxy centre, (2) the molecular hydrogen surface density distribution is strongly correlated with the dust density distribution, and (3) larger hydrogen surface densities appear to be associated with cooler dust than expected when compared to the dust temperatures for the rest of the galaxy, then it stands to reason that those pixels containing this slightly cooler dust should be located towards the centre of the galaxy. We test this hypothesis, highlighting the location on the temperature map of all pixels possessing H$_{2}$ surface densities greater than 100 M$_{\odot}$ pc$^{-2}$ (see Fig. \ref{fig:dustring}). The cooler-than-expected dust is clearly associated with two regions either side of the nucleus.

Finally, we also explored these relations between the dust and gas properties using the \citet{scoville1993} CO(1-0) emission map\footnote{We used Equation 3 in \citet{bolatto2013} to convert the \mbox{CO(1-0)} integrated flux intensity in Jy km s$^{-1}$ into the H$_{2}$ mass (see also e.g., \citealp*{wilson1990}). We adopt the same $D_{L}$ and $X_{CO}$ conversion factor as used in our Equation~\ref{eqn:coconv} ($X_{CO} = 2\, \times$~10$^{20}$~cm$^{-2}$~[K~km~s$^{-1}$]$^{-1}$), giving a H$_{2}$ mass of 2.1~$\,\times\, 10^{9}$~M$_{\odot}$ contained within a thin strip along the major axis, with a mean H$_{2}$ mass surface density of $\Sigma_{\mathrm{H}_{2}} =$~64~M$_{\odot}$ pc$^{-2}$ ranging from 21 to 186~M$_{\odot}$ pc$^{-2}$. We note that we can reproduce the total $M_{\mathrm{H}_{2}}$ value reported in \citet[][5.7~$\times\,10^{9}$~M$_{\odot}$]{scoville1993} if we use their adopted values for $X_{CO}$ and $D_{L}$.}, performing the exact same analysis as described above. Although the CO(1-0) map covers a smaller fraction of the galaxy in the vertical direction compared to the CO(3-2) map (see Fig.~\ref{fig:maps}), we were able to recover the same qualitative correlations between the dust properties and gas components as presented above. In fact, the break in the flattening of the $T_{\mathrm{dust}}$-\sgas \ relation is even more prominent. Our CO(1-0)-based relations only deviated quantitatively from the CO(3-2)-based relations due to a small offset between the two estimates of \shii \ and the fewer number of pixels available for the analysis, which were both expected. Also, due to the lower number of pixels in CO(1-0)-based results, it is ambiguous whether we see a similar bifurcation of the \shii - and \sgas -\sdust \ correlations. Combining all these observations, an interesting picture emerges.

\section{Discussion}\label{sec:discussion}

In summary, whilst we find only a weak correlation between the \hi \ gas content and the dust surface density, we find clear correlations between the dust mass surface density and temperature with the H$_{2}$ and total gas surface densities. These results are interesting for a number of reasons. Firstly, although the CO lines can become optically thick for an edge-on geometry, the strong spatial correlation between the molecular gas surface density and surface density of dust in the majority of the disk of NGC~891 is also observed in face-on galaxies (e.g., \citealp{smith2010}; \citealp{foyle2012}; \citealp{mentuchcooper2012}), indicating that CO emission can also be used as a tracer of the molecular gas distribution in edge-on galaxies. However, our observations hint at the possibility that the CO(3-2) line may become optically thick in the central regions where the column density along the line-of-sight is greatest. Secondly, our gas-to-dust ratios seem to show strong variation in the radial direction along the disk. Similar analyses of the gas and dust using \textit{Herschel} observations of M83 and M51 yielded little or no variation of the gas-to-dust ratio in these face-on systems (\citealp{foyle2012}; \citealp{mentuchcooper2012}), whereas many other galaxy studies have found strong gas-to-dust ratio gradients (e.g., \citealp{munozmateos2009}; \citealp{bendo2010}; \citealp{magrini2011}; \citealp{fritz2012}). Whilst we have yet to fully understand the effects of galaxy structure (e.g. the presence of bars, variations in spirality, etc.) on the gas-to-dust ratio gradients, these features may account for some of the observed radial variations between galaxies. For example, gas tends to be more centrally concentrated in barred galaxies than non-barred galaxies \citep{sakamoto1999}, possibly due to the presence of a bar enhancing gas flows into the central regions (e.g., \citealp{tabatabaei2013b}). Finally, the deviations of individual pixel-by-pixel quantities from these overall correlations in the disk are demonstrably indicative of important morphological features of NGC~891, namely a ring-like feature identified from the relationships between the dust properties and \shii , and evidence of disk asymmetry seen in the spatial distribution of the gas-to-dust ratio. For the remainder of this discussion, we focus our attention on the implications of our results in our understanding of these features of NGC~891.

\subsection{On the molecular ring}\label{sec:molring}

One feature observed in the H$_{2}$-to-dust ratio and cold dust temperature maps suggest the presence of two regions of dense, relatively warmer dust and molecular hydrogen gas located either side of the disk. The regions occupy radial distances between $\sim$ 2 and 5 kpc, which appear almost symmetrical in radial size. Such a feature has already been identified by \citet{israel1999}. In their analysis of the CO and dust emission, they describe the disk structure of NGC~891 via three components comprising a very compact central source / circumnuclear disk (see also \citealp{gb1992}; \citealp{scoville1993}), a `molecular ring' between $R$ = 40\arcsec \ and 120\arcsec \ ($\sim$ 2-6 kpc) and an extended disk detectable out to $R$ = 200\arcsec \ ($\sim$ 9 kpc). This structure in the CO(J=1-0) emission was also noted by \citeauthor{israel1999} to be tightly correlated with the dust traced via the SCUBA 850~$\mu$m emission (see their Fig. 2). The good correlation we observe between the dust surface densities and molecular hydrogen content is thus consistent with these previous results. 

We find that the surface densities of dust and H$_{2}$ are typically high in the molecular ring compared to the rest of the disk distribution. The gas-to-dust ratio clearly peaks on both the NE and SW side of the disk ($\sim$ 3-6 kpc) as the dust surface density declines faster than the H$_{2}$ surface density with increasing radius. The peaks in the H$_{2}$ surface density also correlate spatially with these dust temperature peaks, most likely a consequence of the Kennicutt-Schmidt law relating the H$_{2}$ surface density to the star formation rate (e.g., \citealp{schmidt1959}; \citealp{kennicutt1998}), where star formation subsequently heats the dust. This interpretation seems consistent with the fact our derived dust temperature correlates strongly with the star formation traced via the 24~$\mu$m emission (see Fig.~\ref{fig:dustsourcecomps}, and also the discussion in Section \ref{sec:dustheatinganalysis}). Whilst the molecular ring and the outer extended disk appear to follow linear relationships, a flattening of the relation between the dust temperature and the H$_{2}$ surface density occurs within $R$ = $\pm$40\arcsec \ ($\sim$ 2-3 kpc). Our $T_{\mathrm{dust}}$-\shii \ diagram suggest that two regions of dense, relatively cooler dust and molecular hydrogen gas reservoirs are located on either side of the galaxy center, corresponding roughly to the circumnuclear disk described in \citet{israel1999}. These regions are on average cooler than the temperature we may predict from the overall correlation between $T_{\mathrm{dust}}$ and \shii . The two regions of cooler dust are also evident in the dust temperature map, lying either side of the central temperature peak and between the NE and SW temperature peaks that are coincident with the \shii \ surface density peak. Of course, the  $T_{\mathrm{dust}}$-\shii \ relationship is likely more complex than the linear relationship assumed in this work. Nevertheless, this raises the question of why the dust temperature is on average colder within the inner radius of the molecular ring compared with the rest of the disk. 

FIR observations of giant molecular clouds and filaments have demonstrated that dust temperatures are typically lower in dense molecular clouds than in diffuse regions (see e.g., \citealp{lagache1998}, \citealp{stepnik2003}). The dust and H$_{2}$ are physically associated with star-forming regions and, in dense molecular clouds, dust shields both the H$_{2}$ and CO from dissociation by the ambient interstellar radiation field (ISRF). One explanation for the cooler dust may be that the dust grains in molecular clouds are also opaque to the UV ISRF and so shield the dust embedded deeper within the clouds. An alternative to the possibility of dust shielding in molecular clouds is that there is less dust heating in these regions from star formation. Since both the UV, H$\alpha$ (see e.g., \citealp{kamphuis2007}) and 24~$\mu$m maps (Fig.~\ref{fig:maps}) show a similar dip in emission from the central peak to the two NE and SW maxima (see also the radial profiles in Fig.~\ref{fig:radialprofs}), indicating a relative drop in recent SF between these regions, it is possible that the dust is cooler due to a lack of dust heating compared to the rest of the star-forming disk, particularly in the molecular ring. This is perhaps due to lower star formation rates, or even the aforementioned processes of either self-shielding or molecular gas shielding from heating by the UV interstellar radiation field.  

Perhaps another possible albeit speculative explanation may lie in the process of dust coagulation, where large aggregates are formed with irregular and `fluffy' shapes (see e.g., \citealp{ossenkopf1993}; \citealp{kohler2012}; and references therein). The submillimetre emissivity increases with `fluffiness' (e.g., \citealp{stognienko1995}), whereas the UV to NIR absorptivity remains constant (e.g., \citealp*{bazell1990}), resulting in fluffy aggregates typically having lower equilibrium temperatures compared to compact dust grains (e.g., \citealp*{fogel1998}). Grain-grain coagulation has been invoked to explain the significant dust temperature variations observed in filaments in the Taurus molecular cloud complex (\citealp{stepnik2003}; \citealp{ysard2013}; see also \citealp{paradis2009}). However, given the effects of line-of-sight integration plus the large spatial scales ($\sim$ 0.5 kpc) our observations are probing, we cannot further investigate this scenario. We can only speculate that we are seeing the average effect of dust coagulation or some other process, lowering the observed dust temperature in many unresolved molecular clouds within these regions studied at low resolution. The physical mechanism causing the relationship between the dust temperature and the surface density of molecular hydrogen and total gas remains unknown. Detailed modelling combined with higher resolution data are required to further examine the underlying nature of the correlation.

\subsection{On the disk asymmetry}\label{sec:diskasym}

As we previously mentioned in the introduction, NGC~891 displays a NE-SW asymmetry in the star-forming disk. The NE side has more prominent and extended H$\alpha$ and UV emission than the SW side of the disk (\citealp{dettmar1990}; \citealp{rand1990}; \citealp{kamphuis2007}). Two possible explanations for the nature of this asymmetry have been proposed. \citet{rossa2004} interpret the NE-SW asymmetry as being due to a higher SFR in the northern part of the disk than in the southern part. \citet{kamphuis2007} found that star formation tracers affected by dust attenuation (H$\alpha$, UV) show greater asymmetry compared to those tracers unaffected by dust attenuation (24~$\mu$m, radio continuum). They argue that since the old stellar population and \hi \ gas distribution are fairly symmetric, the asymmetry in H$\alpha$ is most likely caused by attenuation by dust in and above the plane. However, the fact that the small asymmetry is also seen in the radio continuum observations \citep{dahlem1994}, which are not affected by dust attenuation, seems to support the former interpretation. 

\begin{figure}
\begin{center}
\includegraphics[width=0.69\columnwidth]{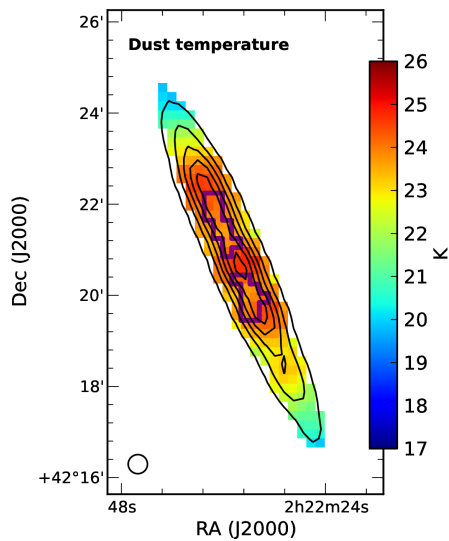}
\end{center}
\vspace{-0.5cm}
\caption[Temperatures of pixels containing large masses of molecular hydrogen]{The map of the dust temperature obtained from fitting a one component, modified blackbody model to the SED of each pixel (\textit{colourscale}, see Fig. \ref{fig:mapsdustprops}). Pixels containing dense regions of molecular hydrogen and cooler-than-average dust are highlighted (\textit{purple contours}). The WISE $F_{12}$/$F_{22}$ flux ratio (\textit{black contours}) follows the dust temperature.}\label{fig:dustring}
\end{figure}

\begin{figure}
\begin{center}
\includegraphics[width=\columnwidth]{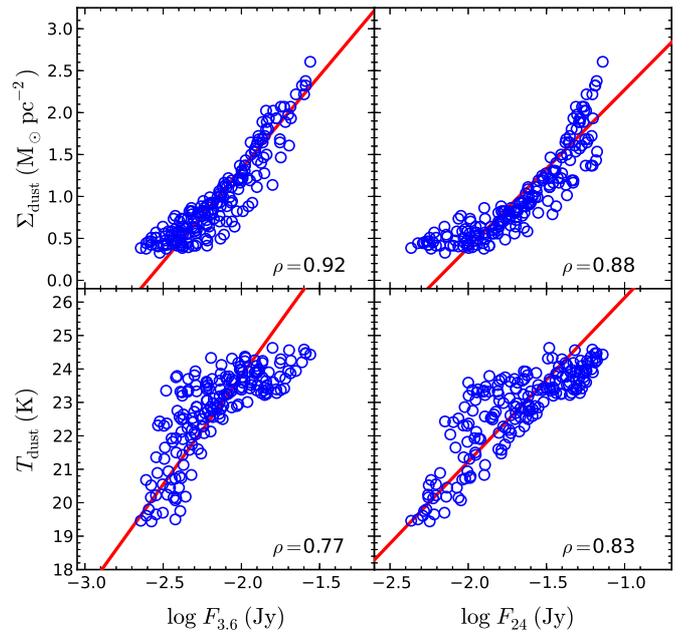}
\end{center}
\vspace{-0.3cm}
\caption[Dust properties versus IRAC and MIPS fluxes]{The pixel-by-pixel relationships between the derived dust properties, \sdust \ (\textit{upper panels}) and $T_{\mathrm{dust}}$ (\textit{lower panels}), and the IRAC~3.6~$\mu$m (\textit{left panels}) and MIPS~24~$\mu$m flux densities (\textit{right panels}). The best linear fits are indicated by the red solid lines. Reported $\rho$ values are the corresponding Pearson correlation coefficients.}\label{fig:dustsourcecomps}
\end{figure}

In order to better understand whether the asymmetry arises due to higher rates of star formation in the north or greater dust obscuration in the south, it is crucial to study the dust surface density and temperature distributions. Firstly, we noted that in the extended disk, beyond the radial extent of the `molecular ring' (i.e. $R >$ 6 kpc), the dust surface density distribution is fairly symmetric between the NE and SW ends of the disk, i.e., both sides contain similar quantities of dust grains available to obscure star-forming regions. Thus, it is unlikely that there is an increase in dust obscuration due to a mere enhancement of the dust distribution in the SW end. However, we cannot completely rule out the scenario due to the presence of a small asymmetry in the distribution. The temperature of the dust in the NE end of the disk is on average warmer (by $\sim$2-3 K) than the dust at corresponding radii on the SW end of the disk (see the dust temperature profile in Fig.~\ref{fig:radialprofs}). This warmer region in the NE is coincident with the peaks in the emission of the H$\alpha$ and UV images (\citealp{kamphuis2007}), and also the peak in the ratio map of 24~$\mu$m to 850~$\mu$m emission (see Fig. 7 in \citealp{whaley2009}; see also our Fig.~\ref{fig:dustsourcecomps}). The latter ratio traces the relative contribution of warm dust associated with star formation and the emission from cold dust. Furthermore, we find that the ISM emission as traced by the WISE $F_{12}/F_{22}$ flux ratio, i.e. the ratio of the emission from polycyclic aromatic hydrocarbons (PAHs) and dust warmed via the UV radiation field and the old stellar population, is also asymmetric (see the black contours in Fig. \ref{fig:dustring}). The contours tracing the ISM are clearly more radially extended on the NE side than the SW side of the disk, and show some overlap with the peak in the cold dust temperature map.

The combined observational evidence suggests a scenario where the warm and cold dust components are being heated by ongoing or recent ($t$ $<$ 10$^{7}$ yr) star formation in the disk, with the asymmetry at various wavelengths arising from an asymmetry in the SFR. The observed asymmetries in the H$_{2}$-to-dust ratio and the total gas-to-dust ratio hint that an enhancement in the SFR may be the result of larger quantities of molecular gas available to fuel star formation in the NE compared to the SW. Furthermore, the substantial \hi \ plume found by \citet{oosterloo2007} suggests the possibility of a continual fuelling of star formation on the NE side, by either the accretion of cooling gas in the halo \citep*{fraternali2008} or the absorption of a small satellite (e.g. \citealp{mouhcine2010}). Recent observations demonstrating solar to supersolar metallicity gas in the halo at 5 kpc from the disk suggests that this halo gas most likely originates from the disk rather than from the accretion of material \citep{bregman2013}. However, the question of whether gas accretion is fuelling ongoing star formation remains.

Finally, we should also consider the insights afforded by the various attempts to interpret NGC~891 using radiative transfer models (e.g., \citealp{kylafis1987}). Smooth, axisymmetric models have so far proved insufficient for accurately fitting the observations of NGC~891. For example, \citet{xilouris1998} found it necessary to take the average optical emission of the northern and southern halves and use an infinitely-long thin disk of stellar emission to reconcile the two sides of NGC~891, whereas smooth models of NGC~891's MIR emission have required extra dust and stellar components to fit the data (\citealp{popescu2000}; \citealp{bianchi2008}). Most recently, \citet{schechtmanrook2012} fitted three-dimensional radiative transfer models including dust clumping and a realistic prescription for spirality to \textit{Hubble Space Telescope} images of NGC~891. Despite some unavoidable degeneracies in their model parameters, the best-fit models determined from a genetic algorithm demonstrated a clear preference for models including spiral structure and clumpy dust (see also the models by \citealp{kamphuis2007}; \citealp{popescu2011}). Taking these results into account, we suspect that the asymmetry likely arises from dust obscuration due to the geometry of the line-of-sight projection of the spiral arms, but we cannot exclude that there is also an enhancement in the star formation rate in the NE part of the disk.

\section{Conclusions} \label{sec:conclusions}

In this work, we investigated the connection between dust and gas in the nearby spiral galaxy NGC~891. High resolution \textit{Herschel} PACS and SPIRE 70, 100, 160, 250, 350, and 500~$\mu$m images are combined with SCUBA 850~$\mu$m observations to trace the far-infrared/submillimetre SED. We fit one-component modified blackbody models to the integrated SED, finding a global dust mass of (8.5 $\pm$ 2.0) $\times$ 10$^{7}$ M$_{\odot}$ and temperature of 23 $\pm$ 2 K, consistent with results from previous far-infrared experiments. We also fit one-component modified blackbody models to pixel-by-pixel SEDs to produce maps of the dust mass and temperature at physical scales of $\sim$ 0.5 kpc. The dust mass distribution shows evidence of several peaks along the disk. The derived dust temperature, which ranges from approximately 17 to 24~K, does not correlate with the dust spatial distribution. Allowing the dust emissivity index to vary, we find an average value of $\beta$ = 1.9 $\pm$ 0.3. We confirm an inverse relation between the dust emissivity spectral index and dust temperature, but do not observe any variation of this relationship with vertical height from the mid-plane of the disk. We adopt a fixed value of $\beta$ = 1.8 for our main analysis comparing the derived dust properties to the gaseous components.

Combining the mapped dust properties with observations of the \hi \ 21~cm line emission and CO(J=3-2) emission to trace the atomic and molecular hydrogen gas, respectively, we examined the correlations between the dust properties with the gaseous components of the ISM. We found strong spatial correlations between the surface mass densities of dust (\sdust) and the molecular hydrogen (\shii) and total gas surface densities (\sgas). We observed no correlation between the dust content and the \hi \ gas content. These observations reveal the presence of regions of dense, cold dust that are coincident with peaks in the gas distribution and appear associated with a molecular ring. Furthermore, the observed asymmetries in the dust temperature, the H$_{2}$-to-dust ratio and the total gas-to-dust ratio hint that an enhancement in the star formation rate may be the result of larger quantities of molecular gas available to fuel star formation in the NE compared to the SW. Whilst we suspect that the asymmetry likely arises from dust obscuration due to the geometry of the line-of-sight projection of the spiral arms, we cannot exclude that there is also an enhancement in the star formation rate in the NE part of the disk.

This study raises several intriguing questions. The observed relationship between the dust temperature and the surface density of the molecular hydrogen suggests that dust may be cooler in dense regions of dust and molecular hydrogen, perhaps due to the processes of either self-shielding or molecular gas shielding from heating by the UV interstellar radiation field, or dust coagulation. However, whilst these processes have support from observations at high spatial resolution in nearby molecular clouds, it is unclear whether our lower resolution dataset is capable of uncovering such a correlation between the dust temperature and molecular gas density observed on these large spatial scales. This remains an open issue, which may be resolved by a detailed analysis and modelling of the $T_{\mathrm{dust}}$-\shii \ relation based on larger samples of edge-on galaxies, e.g. \textit{HERschel} Observations of Edge-on Spirals \citep[\textit{HER}OES, ][]{verstappen2013} and the New \textit{HErschel} Multi-wavelength Extragalactic Survey of Edge-on Spirals \citep[NHEMESES, ][]{holwerda2012}. Furthermore, it is still an open question whether molecule formation is mainly driven by hydrostatic pressure or UV radiation shielding over different spatial scales and over a large range of metallicities (see e.g., \citealp{fumagalli2010} and references therein). Upcoming facilities such as the Atacama Large Millimeter Array and the Square Kilometer Array should shed some light on this problem. These interferometers will allow the mapping of atomic and molecular gas and dust at high resolution, potentially enabling the study of how the relative quantities of these ISM components vary with metallicity and photodissociating UV radiation intensity in different environments. A formalisation of the processes governing the molecular content in galaxies will undoubtedly aid future cosmological simulations of galaxy formation and evolution. 

\begin{acknowledgements}
We thank an anonymous referee for comments and suggestions, which improved the quality of the paper. TMH wishes to thank T. Oosterloo for kindly sharing the {\sc FITS} files of the \hi~data, and F. Israel and N. Scoville for their CO observations, all of which were crucial for this work. TMH, MB and JF gratefully acknowledge the financial support from the Belgian Science Policy Office (BELSPO) in the frame of the PRODEX project C90370 (Herschel-PACS Guaranteed Time and Open Time Programs: Science Exploitation). IDL and FA are postdoctoral and doctoral fellows of the Flemish Fund for Scientific Research (FWO-Vlaanderen), respectively. MB and SV also acknowledge the financial support of the same Flemish Fund for Scientific Research. PACS has been developed by a consortium of institutes led by MPE (Germany) and including UVIE (Austria); KU Leuven, CSL, IMEC (Belgium); CEA, LAM (France); MPIA (Germany); INAF-IFSI/OAA/OAP/OAT, LENS, SISSA (Italy); IAC (Spain). This development has been supported by the funding agencies BMVIT (Austria), ESA-PRODEX (Belgium), CEA/CNES (France), DLR (Germany), ASI/INAF (Italy), and CICYT/MCYT (Spain). SPIRE has been developed by a consortium of institutes led by Cardiff University (UK) and including Univ. Lethbridge (Canada); NAOC (China); CEA, LAM (France); IFSI, Univ. Padua (Italy); IAC (Spain); Stockholm Observatory (Sweden); Imperial College London, RAL, UCL-MSSL, UKATC, Univ. Sussex (UK); and Caltech, JPL, NHSC, Univ. Colorado (USA). This development has been supported by national funding agencies: CSA (Canada); NAOC (China); CEA, CNES, CNRS (France); ASI (Italy); MCINN (Spain); SNSB (Sweden); STFC, UKSA (UK); and NASA (USA). This research has made use of the NASA/IPAC Extragalactic Database (NED) which is operated by the Jet Propulsion Laboratory, California Institute of Technology, under contract with the NASA (USA). This research made use of Montage, funded by the NASA (USA) Earth Science Technology Office, Computation Technologies Project, under Cooperative Agreement Number NCC5-626 between NASA and Caltech, and maintained by the NASA/IPAC Infrared Science Archive. This publication makes use of data products from the Wide-field Infrared Survey Explorer, which is a joint project of the University of California, Los Angeles, and the JPL/Caltech, funded by NASA (USA).

\end{acknowledgements}

\bibliography{aa23245-13}

\end{document}